\documentclass[12pt,english]{iopart}

\usepackage{amssymb}
\usepackage{graphicx}
\usepackage{dcolumn}
\usepackage{bm}

\usepackage{braket}
\usepackage[T1]{fontenc}	
\usepackage[utf8x]{inputenc}
\usepackage[english]{babel} 				%

\usepackage{cite}
\usepackage{color}
\usepackage{dsfont}

\usepackage{overpic}

\makeatletter
\@ifundefined{textcolor}{}
{%
 \definecolor{BLACK}{gray}{0}
 \definecolor{WHITE}{gray}{1}
 \definecolor{RED}{rgb}{1,0,0}
 \definecolor{GREEN}{rgb}{0,1,0}
 \definecolor{BLUE}{rgb}{0,0,1}
 \definecolor{CYAN}{cmyk}{1,0,0,0}
 \definecolor{MAGENTA}{cmyk}{0,1,0,0}
 \definecolor{YELLOW}{cmyk}{0,0,1,0}
 }

\makeatother

\begin{document}

\title[Single Atom Collisionally Coupled to an Ultracold Finite
Bosonic Ensemble]{Correlated Quantum Dynamics of a Single Atom Collisionally
Coupled to an Ultracold Finite Bosonic Ensemble}

\author{Sven Kr\"onke$^{1,2}$, Johannes Kn\"orzer$^{1,2}$ and Peter
Schmelcher$^{2,3}$}
 \address{$^1$These authors contributed equally to this work.}
 \address{$^2$Center for Optical Quantum Technologies, University of Hamburg,
Luruper Chaussee 149, 22761 Hamburg, Germany}
\address{$^3$ The Hamburg Centre for Ultrafast Imaging, University of Hamburg,
Luruper Chaussee 149,
22761 Hamburg, Germany}

\ead{\mailto{sven.kroenke@physnet.uni-hamburg.de}, 
\mailto{johannes.knoerzer@physnet.uni-hamburg.de}, 
\mailto{peter.schmelcher@physnet.uni-hamburg.de}}

\date{\today}

\begin{abstract}
We explore the correlated quantum dynamics of a single atom, regarded as an
open system, with a spatio-temporally localized coupling to a finite bosonic
environment. The single atom, initially 
prepared in a coherent state of low energy, oscillates in a one-dimensional
harmonic trap and thereby periodically penetrates an interacting ensemble
of $N_A$ bosons, held in a displaced trap. 
We show that the inter-species energy transfer accelerates with increasing 
$N_A$ and becomes less
complete at the same time. System-environment correlations prove
to be significant except for times when the excess energy distribution
among the subsystems is highly imbalanced. These correlations result in
incoherent energy transfer processes, which accelerate the early energy
donation of the single atom and stochastically 
favour certain energy transfer channels
depending on the instantaneous direction of transfer. 
Concerning the subsystem states,
the energy transfer is mediated by non-coherent states of the single atom and 
manifests itself in singlet and
doublet excitations in the
finite bosonic environment. 
These comprehensive insights into the non-equilibrium quantum 
dynamics of an open system are gained by {\it ab-initio} simulations
of the total system with the recently developed
Multi-Layer Multi-Configuration Time-Dependent Hartree Method
for Bosons.
\end{abstract}


\maketitle

\section{Introduction}\label{sec:intro}

Many physically relevant quantum systems are in fact open.
Intriguing effects in
e.g. condensed matter 
physics \cite{quantum_diss_systems_Weiss}, quantum optics
\cite{open_sys_approach_qo}, molecules or light harvesting
complexes \cite{vendrell_proton_2008,ritschel_efficient_2011,roden_probability_2015}
are
intimately related to environmentally induced dissipation and decoherence and
thus
require a careful treatment beyond the unitary dynamics of the time-dependent
Schr\"odinger equation.  As
one can often neither experimentally control nor theoretically
describe all
the environmental degrees of freedom, a variety of theoretical methods has
been developed for an effective description of the reduced dynamics of the open
quantum system of interest over the last decades
\cite{Breuer_theory_open_quantum_systems}. 

Due to their high degree of controllability and, in particular, isolatedness
\cite{many_body_physics_ultrac_atoms_zwerger_dalibard_bloch},
ensembles of ultracold atoms or ions serve as ideal systems in order to
systematically
study the dynamics of open quantum systems allowing for various
perspectives on this subject \cite{muller_engineered_2012}. Open quantum system
dynamics has been implemented by
digital quantum simulators
\cite{barreiro_open-system_2011,schindler_quantum_2013} or by partitioning an
ultracold atomic ensemble into carefully
coupled, distinguishable subsystems
\cite{scelle_motional_2013}. As a matter of fact, many impurity problems can
also be viewed as open quantum system settings: enormous experimental progress
such as
\cite{
bakr_quantum_2009,
sherson_single-atom-resolved_2010,
weitenberg_single-spin_2011,
deterministic_preparation_of_tunable_few-fermion_system_selim_Science2011}
allows to prepare a few impurities or even a single one in an ensemble of
atoms in
order to study thermalization and atom loss mechanisms
\cite{spethmann_dynamics_2012}, transport and polaron physics
\cite{palzer_quantum_2009,johnson_impurity_2011,fukuhara_quantum_2013} or the
damping of the breathing mode \cite{catani_quantum_2012,peotta_quantum_2013-1}.
Such implementations of open quantum systems offer the unique flexibility to
tune both the character and strength of the system-environment coupling
\cite{catani_quantum_2012} and the environmental properties
\cite{fukuhara_quantum_2013}. Moreover, dissipation and environment engineering 
can also be employed for controlling many-body
dynamics \cite{barontini_controlling_2013}, state preparation
\cite{diehl_quantum_2008} as well as quantum computation
\cite{verstraete_quantum_2009,pastawski_quantum_2011}.

In this work, we theoretically study the open
quantum system dynamics of a single atom with a weak
spatio-temporally localized coupling to a finite bosonic environment,
focusing in particular on the characterization of the inter-species energy
transfer processes.
By considering a binary mixture of neutral atoms interacting via contact
interaction both within and between the species, the local character of the
coupling is realized. In order to localize the coupling also in time,
species-selective one-dimensional trapping potentials as well as a particular
initial
condition are considered: the single
atom is initially displaced from the centre of its harmonic
trap, i.e. resides in a coherent state, while the bosonic ensemble is prepared
in the ground state
of a harmonic trapping potential being shifted from the trap of the single atom.
Thereby, both subsystems couple only during a certain phase 
 of the 
single atom oscillation and the effective coupling strength becomes strongly 
dependent on the instantaneous subsystem states. In a
different context, this kind of coupling has already been realized effectively
experimentally \cite{harter_single_2012}. 
The basic ingredient of such a coupling, a single binary collision,
can result in entanglement between the collision partners 
\cite{coll_gate_Zoller,mack_dynamics_2002,hutton_entangle_by_scattering,
benedict_time_2012}, which has been shown to depend on the details such as
the scattering phase shift, the mass ratio as well as the relative momentum of 
the atoms \cite{law_entanglement_2004}. As a consequence,
significant correlations between the single atom and the finite bosonic
environment can occur after many collisions despite of possibly weak
interactions
 (cf. also
\cite{mack_dynamics_2002,sowinski_dynamics_2010,
Gardiner_PRA2013} in this context),
undermining a mean-field
approximation for the two species on a longer time scale.
We note that the scenario under consideration 
might seem to be reminiscent of the quantum Newton's cradle
\cite{kinoshita_quantum_2006,ganahl_quantum_2013,franzosi_newtons_2014}. Yet instead of
investigating the (absence of) thermalization in a closed system of
indistinguishable constituents,  we are concerned with unravelling the interplay
of energy transfer between distinguishable subsystems and the
emergence of correlations when systematically increasing the number of
environmental degrees of freedom $N_A$.

Rather than simulating the reduced dynamics of the single atom only, we employ
the recently developed {\it ab-initio} Multi-Layer Multi-Configuration
Time-Dependent Hartree Method for Bosons
(ML-MCTDHB) \cite{kronke_non-equilibrium_2013,cao_multi-layer_2013} for
obtaining the
non-equilibrium quantum dynamics of the whole system for various numbers of
environmental degrees of freedom. Such a closed system perspective on an open
quantum system problem gives the unique opportunity to
investigate not only the dynamics of the
open system but also its impact on the environment and, moreover, to
systematically uncover correlations between the two subsystems
(cf. also e.g. \cite{sys_env_corr_non_mark_ansgar,env_collec_reaction_coord}).
Full numerical simulations are supplemented with analytical and perturbative
considerations.

This paper is organized as follows: In sect. \ref{sec:setup}, we
introduce our setup and discuss a possible experimental implementation.
Moreover, we argue how
the inter-species energy transfer channels can be controlled by
appropriately tuning the separation of the involved species-selective
traps. The subsequent sect. \ref{sec:energy} is devoted to the energy transfer
dynamics between the single
atom and the finite bosonic environment.
Here, we
show that the energy transfer between the subsystems is accelerated with
increasing $N_A$ due to a level splitting of the involved excited many-body
states. On top of it, the relative amount of the total energy being
exchanged between the subsystems is reduced when increasing $N_A$. 
After an initial slip
the energy distribution among the subsystems consequently fluctuates about quite
a balanced one for larger $N_A$.
In sect.
\ref{sec:state_analysis}, we investigate how the 
subsystem states are affected by the system-environment coupling: By
inspecting the Husimi phase space distribution of the single atom, we conclude
that the energy transfer is mediated by non-coherent states manifesting
themselves in a drastic deformation of the phase space distribution during
relatively short time slots. Concerning the environment,
depletion oscillations are observed, whose amplitude decreases with increasing
$N_A$ when fixing the initial displacement of the single atom.
In sect. \ref{sec:corr_subsystems}, we unravel the oscillatory emergence and
decay of inter-species correlations showing that whenever the excess energy
distribution among the subsystems is highly imbalanced, correlations have to be
strongly suppressed. The maximal attained correlations turn out to be
independent of the considered number of environmental degrees of freedom $N_A$.
By analysing the subsystem energy distribution among the so-called natural
orbitals, we show how inter-species correlations result in
incoherent energy transfer processes, which accelerate the early energy
donation of the single atom. Most importantly, 
we ultimately uncover the interplay between subsystem excitations and
correlations by means of a Fock space excitation analysis in sect. 
\ref{sec:exc_fock},
characterizing the environmental excitations
and unravelling correlations between subsystem excitations with a
tailored correlation measure.
Thereby, we can show that inter-species
correlations (dis-)favour certain energy transfer channels depending on the
instantaneous direction of transfer in general.
Finally, we conclude and give an outlook in sect. \ref{sec:concl}.

\section{Setup and initial conditions}\label{sec:setup}

The subject of this work is a bipartite system confined to a single spatial
dimension and consisting of two atomic species:
a single atom is collisionally coupled to a finite bosonic environment (cf.
fig. \ref{fig:setup} (c)).
We model the intra- and inter-species short-range interactions with a contact
potential as it is usually done in the ultracold
s-wave scattering limit \cite{Pethick2002}.
In addition, we assume two external harmonic potentials for a species-selective
trapping of the atoms.
In the following, we adopt a shorthand notation labelling the bosonic
environment with $A$ and the single atom with $B$.
Throughout this work, all quantities are given in terms of natural harmonic
oscillator units of $B$, which
base on the mass of the single atom $m_B$ and its trapping frequency
$\omega_B$.
Lengths, energies and times are thus given in terms of 
$\sqrt{\hbar/(m_B\omega_B)}$, 
$\hbar \omega_B$ and $\omega_B^{-1}$, respectively.
The Hamiltonian of the composite
system takes the form
\begin{equation}\label{eq:hamilt_sep}
\hat H = \hat H_A + \hat H_B + \hat H_{AB}
\end{equation}
with the Hamiltonian of the environment $\hat H_A$, the open system $\hat H_B$ 
and the inter-subsystem coupling $\hat H_{AB}$:
\begin{eqnarray}\label{eq:hamilt_parts}
 \hat H_{A} &= \sum_{i=1}^{N_A} \left ( \frac{1}{2 \beta} ({\hat p^A_i})^2 +
\frac{\alpha^2 \beta}{2} (\hat x^A_i + R)^2 \right ) + g_A \sum_{1 \leq i < j \leq
N_A} \delta(\hat x^A_i-\hat x^A_j), \nonumber \\
   \hat H_{B} &= \frac{1}{2} ({\hat p^B})^2 + \frac{1}{2} (\hat x^B)^2, \\
   \hat H_{AB} &= g_{AB} \sum_{k=1}^{N_A} \delta(\hat x^{B} - \hat x^A_k). 
   \nonumber
\end{eqnarray}
The dimensionless representation of the Hamiltonian involves 
the mass ratio $\beta = m_A/m_B$ and the frequency ratio $\alpha =
\omega_A / \omega_B$. $R>0$ denotes the separation between the traps 
and the intra- and inter-species interaction strengths are given by $g_A$ and
$g_{AB}$, respectively.
We emphasize that the variety of parameters specifying the Hamiltonian
(\ref{eq:hamilt_sep}) accounts for the
versatility of the considered system, which to explore in its complexity goes
far beyond this work.
Rather than this, we fix $g_A, g_{AB}$ to experimentally feasible values by
considering two $^{87}$Rb
hyperfine states of comparable intra- and inter-species scattering lengths,
implying $\beta = 1$.
Assuming $\omega_B = 10\,{\rm Hz}$ and a transverse to longitudinal trapping
frequency ratio of about $68$ for both species, we are left with $g_A =
g_{AB} = 0.08$ for the one-dimensional coupling constants after dimensional
reduction \cite{Olshanii_PRL98_quasi_1d_scattering,
Pitaevskii_Stringari_Bose-Einstein_Condensation2003}.

\definecolor{ao}{rgb}{0.0, 0.5, 0.0}

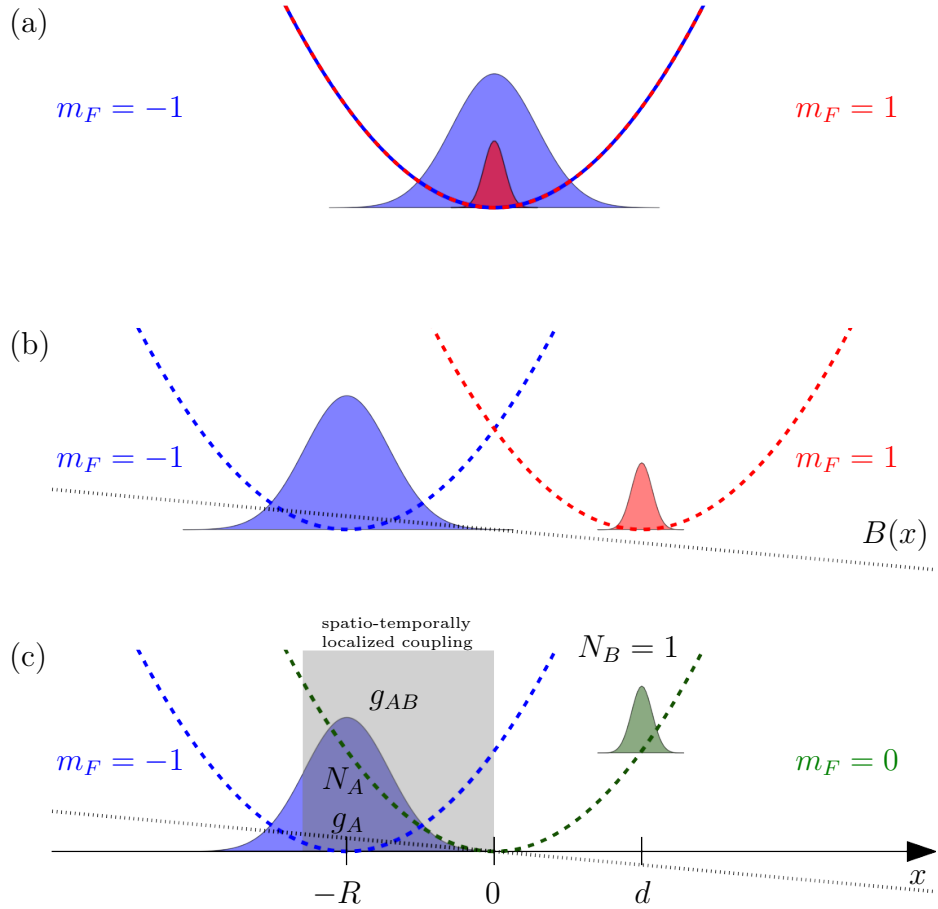
\begin{figure}[t!]
\centering
\begin{overpic}[width=0.8\textwidth]
{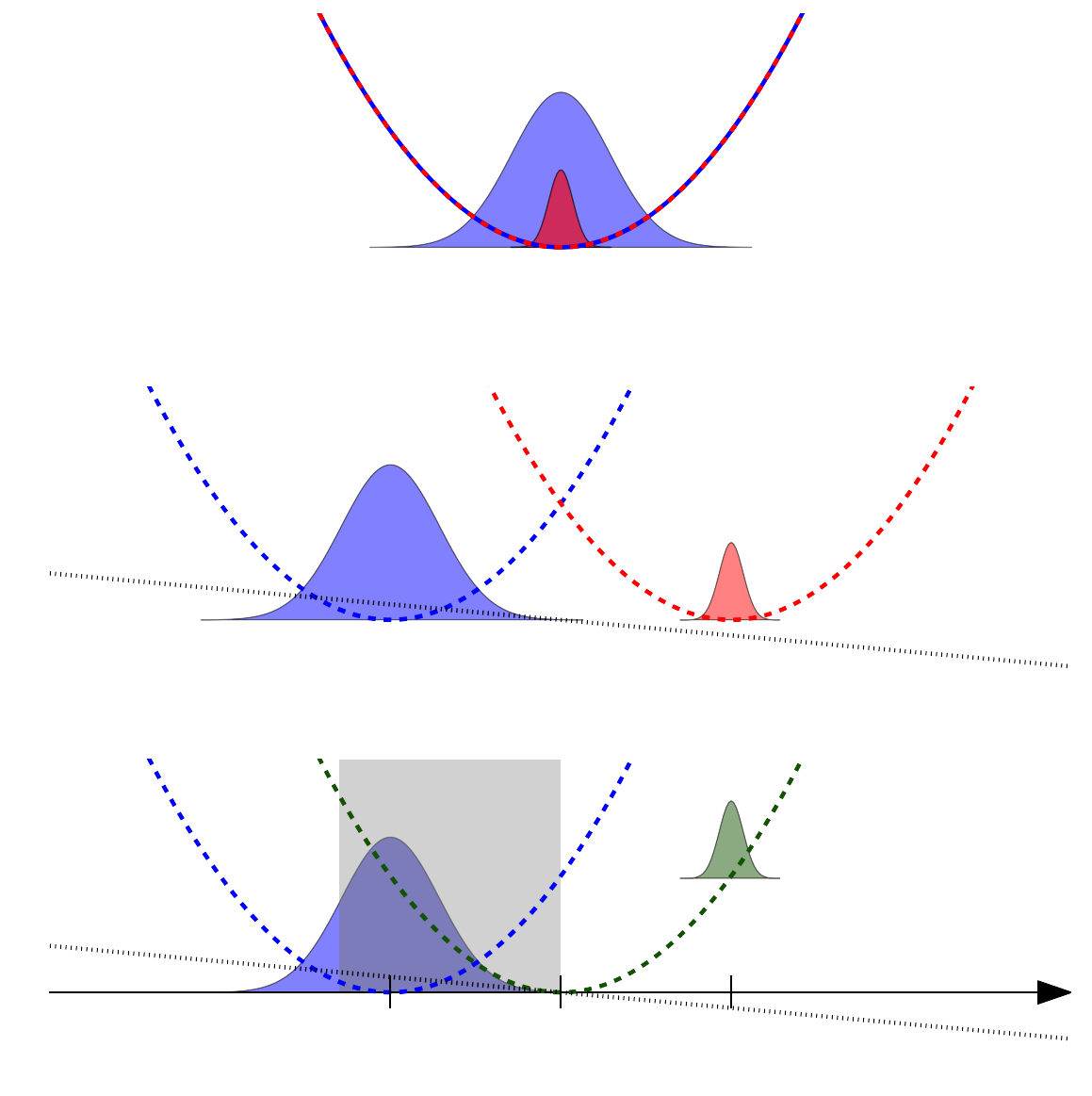}
\put(0,96){(a)}
\put(0,62){(b)}
\put(0,29){(c)}
\put(5,87){\color{blue}{$m_F = -1$}}
\put(5,50){\color{blue}{$m_F = -1$}}
\put(5,18){\color{blue}{$m_F = -1$}}

\put(83,87){\color{red}{$m_F = 1$}}
\put(83,50){\color{red}{$m_F = 1$}}
\put(83,18){\color{ao}{$m_F = 0$}}
\put(90,42){$B(x)$}
\put(95,6){$x$}
\put(33,33){\tiny {spatio-temporally}}
\put(33,31){\tiny {localized coupling}}
\put(38,25){$g_{AB}$}
\put(33,16){$N_A$}
\put(60,30){$N_B = 1$}
\put(34,12){$g_A$}
\put(32,4){$-R$}
\put(50.25,4){$0$}
\put(66,4){$d$}
\end{overpic}
\caption[Caption_2]{Sketch of a possible three-step realization of the
bipartite system with the single atom initially prepared in a coherent state.
(a) An optical dipole trap holds $N_A$ atoms ($m_F = -1$) and a single impurity
atom ($m_F = +1$). (b) Application of an external magnetic field gradient
yields a spatial separation of the ensemble and the impurity. (c) A RF field
drives the $m_F = 1$ to $m_F = 0$ transition and initializes the single atom in
a displaced ground state.}
\label{fig:setup}
\end{figure}

We initialize $B$ in a coherent state $\ket{z}$:
\begin{equation}\label{eq:coherent_state}
\ket{z} = e^{-|z|^2/2} \sum_{n=0}^{\infty} \frac{z^n}{\sqrt{n!}}\ket{u^B_n}
\end{equation}
where $\ket{u^\sigma_n}$ denotes the $n^{th}$ harmonic oscillator (HO)
eigenfunction
of the species $\sigma = A, B$
with energy $E^A_n =\alpha(1/2 + n)$, $E^B_n = E^A_n/
\alpha$, respectively. We choose $z = d / \sqrt{2} > 0$
such that $\ket{z}$ equals the $\hat H_B$ ground state displaced by a 
distance $d$ to the right from the $B$ species trap centre.
The environment, $A$, in turn is prepared in the ground state
of $\hat H_A$. In order to realize the desired spatio-temporally localized
coupling, whose impact on the inter-species energy transfer and emergence of
correlations shall
be investigated, we require $R$ and $d$ to be such that there is (i) no
inter-species overlap at $t = 0$ and
(ii) finite overlap after half of a $B$ atom oscillation period, i.e. at
$t\approx\pi$.
For more details on the initial state preparation and
subsequent propagation with our \textit{ab-initio} ML-MCTDHB
method, see \ref{app:method} and \ref{app:init_converg}.

In order to characterize the spatio-temporally localized
coupling, we express $\hat H_{AB}$ in terms of the HO eigenbasis $\left\{
|u^\sigma_n\rangle \right\}$,
which serves as a natural choice for the low-energy and weak-coupling regime we
are interested in,
\begin{equation}\label{eq:hamilt_ab_repr2}
 \hat H_{AB} = g_{AB} \sum_{i,j,p,q} v_{ijpq} {\hat a}^\dagger_i {\hat a}_p 
 \otimes |u^B_j\rangle\!\langle u^B_q|,
\end{equation}
 with $\hat a^{(\dagger)}_i$ denoting the bosonic annihilation (creation)
operator corresponding to $|u^A_i\rangle$ and the interaction-matrix elements
given by
\begin{eqnarray}\label{eq:v_ijpq}\fl
 v_{ijpq} &= \braket{u^A_i u^B_j | \delta(\hat x^A-\hat x^B) | u^A_p u^B_q} =
\int_{-\infty}^{+\infty} dx \ [u_i^A(x) u^B_j(x)]^* u^A_p(x) u^B_q(x),
\end{eqnarray}
which obviously depends on $\alpha$ and $R$.
Eventually, we are interested in rather small displacements $d$ for which the
matrix element $v_{1001}$ is of major importance for the dynamics.
For e.g. $\alpha = 1$, we find that $v_{1001} = (1-R^2) \exp (-R^2/2) / (2
\sqrt{2\pi})$.
Hence, the scattering channels "$\ket{u_0^A}\ket{u_1^B} \leftrightarrow
\ket{u_1^A}\ket{u_0^B}$" are completely suppressed at $R = 1$.
In order to avoid artefacts of such selection rules, we choose $R=1.2$,
for which the relevant low-energy scattering channels are not suppressed.

Although we have fixed some of the parameters, the system still features a high
sensitivity to $\alpha$, $d$, $N_A$ and thus controllability.
For $N_A = 1$, analytical expressions for the energy spectrum and the wave
functions
can be worked out and trap-induced shape resonances between molecular and trap
states are detectable \cite{krych_displaced_traps,busch_nonlocality}.
Since this two-body problem has already been treated in detail, we shall rather
focus on the impact of the environment size, considering $N_A = 2, ..., 10$.

Finally, we briefly comment on an experimental realization of this open quantum
system problem. 
Having loaded an ensemble of ${|F=1,m_F=-1\rangle}$ polarized 
$^{87}$Rb atoms
in an optical dipole trap with a deep transverse optical lattice, the
site-selective 
spin flip technique based on an additional longitudinal pinning lattice
\cite{weitenberg_single-spin_2011} or the doping technique in
\cite{spethmann_dynamics_2012}
allows for creating a single $|F=1,m_F=1\rangle$ impurity
(fig. \ref{fig:setup} (a)). By adiabatically
ramping up a longitudinal magnetic field gradient,
the species are spatially displaced in opposite directions
(fig. \ref{fig:setup} (b)). Then the $m_F=1$ to
$m_F=0$ transition can
be selectively addressed by a RF field using the quadratic Zeeman effect such
that the single $B$ atom ($m_F=0$) is
initialized in a displaced coherent state, realizing $d=R$ and $\alpha=1$
(fig. \ref{fig:setup} (c)).
The energy of the $B$ atom
can be inferred from its oscillation turning points via
{\it in situ} density
\cite{bakr_quantum_2009,
sherson_single-atom-resolved_2010} or tunnelling measurements
\cite{fermionizations_of_2_distinguishable_fermions_ZuernSelim_PRL2012}.

\section{Energy transfer}\label{sec:energy}

In order to understand which kind of energy transfer processes 
between the species are feasible with the spatio-temporally localized coupling
and how efficient these are,
we examine the energies of the subsystems, identified with 
$\braket{\hat H_{A}}_t$ and $\braket{\hat H_{B}}_t$, separately. 
Since the interaction energy
$\braket{\hat H_{AB}}_t$ cannot be attributed to the energy content of a
single species, the aforementioned identification is problematic, in general.
However, the
spatio-temporally localized coupling always allows to find times during a $B$
atom oscillation at which the inter-species interaction is essentially
negligible so that $\braket{\hat H_{A}}_t$ and
$\braket{\hat H_{B}}_t$ then measure how the energy is distributed
among the open system and its environment. Because of the weak
and spatio-temporally localized inter-species coupling, the short-time dynamics
taking place on the time-scale of the free oscillation period of the $B$ atom, 
i.e. $T=2\pi$, is clearly separated from the long-time dynamics of the energy 
transfer between the two species. When quantifying times, we will use the term
``$B$ oscillations'' 
always with reference to free harmonic oscillations of the $B$ atom.
Due to this time-scale separation, we present the expectation value of any
physical observable $\hat{O}$ as locally time-averaged over one free
$B$ oscillation period,
\begin{equation}
\label{eq:forward_averaging}
\bar{O}(t) = \frac{1}{2\pi} \int_t^{t+2\pi} d \tau
\braket{\Psi_t|\hat{O}|\Psi_t}.
\end{equation}
Thereby, we average out fluctuations on the short time-scale of a $B$
oscillation, giving a clearer view on the long-time dynamics.
The fast short-time
dynamics is captured by the variance of the local time-average,

\begin{equation}
\label{eq:forward_averaging_var}
{\rm var}(\hat{O};t) = \frac{1}{2\pi} \int_t^{t+2\pi} d \tau
\braket{\Psi_\tau|\hat{O}^2|\Psi_\tau}-\bar{O}_t^2.
\end{equation}
\noindent In some cases, it turns out to be insightful to encode this
information in the figures by accompanying the lines corresponding to
$\bar{O}(t)$ with a shaded area indicating the standard deviation.

We prepare the system such that the two species have no spatial overlap at $t =
0$. Therefore, the initial energy of the single atom reads
\begin{equation}\label{eq:e_b}
 \langle \hat H_B\rangle_{t=0} = \frac{d^2+1}{2}.
\end{equation}
By adjusting the displacement $d$, we can thus vary the amount of deposited
excess energy $\varepsilon \equiv d^2 / 2$, which is available
for energy transfer processes between the subsystems. These
energy transfer processes are captured in terms of the
intra-species excess energies $\varepsilon^\sigma_t=
\langle \hat H_\sigma\rangle_t-E^\sigma_{\rm gs}$ with $E^\sigma_{\rm gs}$
denoting the ground state energy of $\hat H_\sigma$, $\sigma=A,B$. We employ
the normalized excess energy of $B$, $\Delta^B_t=\varepsilon^B_t/\varepsilon$,
in order to measure how balanced the deposited energy is distributed among
the subsystems: At instants for which
$\langle \hat H_{AB}\rangle_t$ is negligible, one obviously finds 
$\varepsilon^A_t/\varepsilon=1-\Delta^B_t$. Thus $\Delta^B_t\approx 1 (0)$
implies then
that almost all excess energy is stored in the $B$ ($A$) species, corresponding
to a maximally imbalanced excess energy distribution, while 
$\Delta^B_t\approx 0.5$ refers to a balanced distribution. Due to the
initial imbalance $\Delta^B_0=1$, the $B$ atom will first donate energy
to its environment implying an overall decrease of $\Delta^B_t$ until 
$\Delta^B_t$ reaches a minimum. We call the energy donation of $B$ the more
efficient the closer to zero this minimum is. Afterwards, the $B$ atom will 
generically accept energy from its environment, resulting in an
overall increase of $\Delta^B_t$ until it reaches a maximum. We call such an
energy transfer cycle the more complete the closer to unity this maximum is.

First of all, we investigate how the geometric properties of the trap and the 
initial excess energy influence the energy donation
by inspecting $\max_t (1 - \Delta^B_t)$ for various values of $d\in [ 0.5, 3 ]$
and $\alpha\in [2/3,4/3]$ and for a fixed, reasonably long propagation         
time.           
While the maximal fraction of excess energy transferred to the A species is
rather independent of d, we observe a sharp resonance at $\alpha=1$
(plot not shown).
As we will show (cf. \ref{app:proof_disentanglement}), intriguing open-quantum
system dynamics with a significant
impact of inter-species correlations can only take place for excess energies
$\varepsilon$ being at least of the order of the excitation gap of $\hat H_A$,
which equals $\alpha$
when neglecting the intra-species interactions. In this work, we thus
concentrate on the
resonant case $\alpha=1$ with $\varepsilon \approx
\alpha$ so that
not too many scattering channels are energetically open, which paves the way for
a thorough
understanding of the dynamics. If not stated otherwise, $d=1.5$ is consequently
assumed leaving us with $N_A$ as the only free parameter.

\begin{figure}[t!]
\centering
\includegraphics[width=0.8 \textwidth]{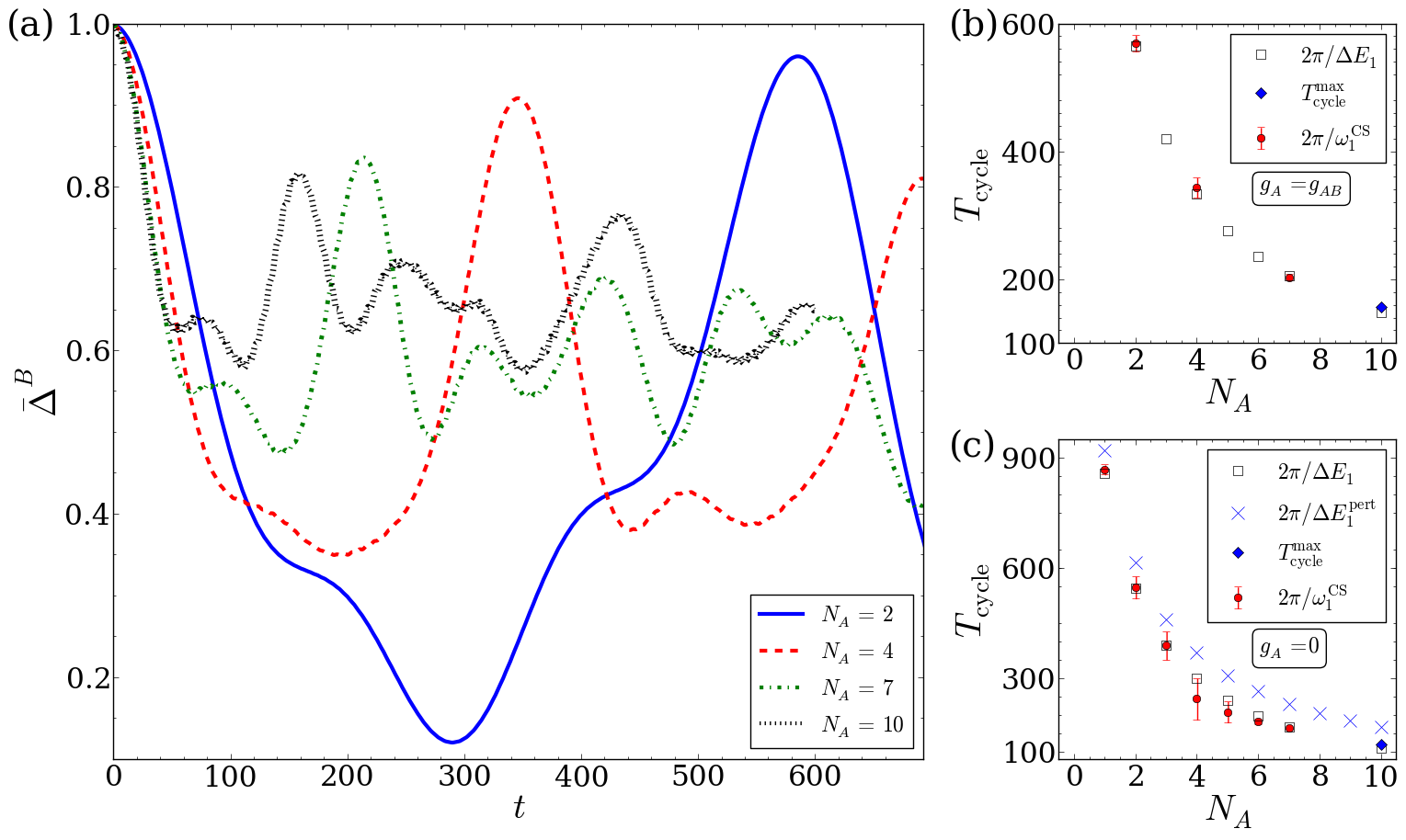}
\caption[Caption_3]{(a) Evolution of the excess energy imbalance $\bar
\Delta^B(t)$ for various $N_A$.
(b) Time-scale $T_{\rm cycle}=2\pi/\omega_1^{\rm CS}$ of an energy transfer
cycle obtained by
compressed sensing of $\langle \hat H_{B}\rangle_t$ simulation data (full circles,
error bars indicate the standard deviation related to the corresponding peak
width in the
frequency spectrum). For $N_A=10$, 
we provide the instant $T_{\rm cycle}^{\rm max}$ of the first significant $\bar
\Delta^B(t)$ maximum besides $t=0$ (blue diamond) instead of $2\pi/\omega_1^{\rm
CS}$.
Squares denote the time-scale $2\pi/\Delta E_1$
induced by the
level spacing $\Delta E_1$ of the first excited manifold calculated with
ML-MCTDHB. (c) Same as (b) but for $g_A=0$. Crosses refer to
the time-scale $2\pi/\Delta E_1^{\rm pert}$
with $\Delta E_1^{\rm pert}$ denoting the first order
perturbation theory approximation for $\Delta E_1$.
Other parameters: $\alpha=1$, $d=1.5$.
All quantities in HO units of $\hat H_B$.}
\label{fig:energies}
\end{figure}

Fig. \ref{fig:energies} shows the normalized excess energy $\bar{\Delta}^B(t)$
for $N_A = 2, 4, 7, 10$.
As a starting point, we discuss the case $N_A = 2$. The initial excess energy
distribution is maximally imbalanced, i.e. $\Delta^B_{t=0} = 1$.
As $B$ donates energy to its environment, $\bar{\Delta}^B(t)$ monotonously
decreases until reaching a global minimum at $t \approx 289$,
implying a directed energy transfer from $B$ to $A$.
The energy transfer decelerates at $t \approx 145$ and reaccelerates at $t
\approx 190$.
When $\bar{\Delta}^B(t)$ attains its minimum at $t \approx 289$, the reverse
energy transfer process is initiated
and $B$ accepts the energy that was donated to $A$ in the first place. 
Decelerated energy transfer can also be observed
between $t \approx 395$ and $t \approx 440$. At $t \approx 584$,
$\bar{\Delta}^B(t)$ attains a local maximum and the excess energy is almost
completely
restored in $B$ ($\bar{H}_B(t=584) \gtrsim 0.97 \braket{\hat H_B}(t=0)$).

Now we turn to the impact of $N_A$ on $\bar{\Delta}^B(t)$.
The oscillatory evolution of $\bar{\Delta}^B(t)$ is similar for $N_A = 2$ and
$N_A = 4, 7, 10$. For example, we identify the minimum of
$\bar{\Delta}^B(t)$ for $N_A = 2$ at $t \approx 289$ with the minimum of
$\bar{\Delta}^B(t)$ for $N_A = 4$ at $t \approx 195$.
Analogously, the maximum of $\bar{\Delta}^B(t)$ for $N_A = 2$ at $t \approx 584$
indicating that $B$ accepted much of the excess energy $\varepsilon$
from the environment corresponds to the maximum of $\bar{\Delta}^B(t)$ for 
$N_A=4$ at $t \approx 345$.
Already after a few $B$ oscillations, $\bar{\Delta}^B(t)$ is considerably
smaller
for higher $N_A$ than for $N_A = 2$.
Thus, all results shown in fig. \ref{fig:energies} indicate that the overall
energy transfer process accelerates with increasing $N_A$.
Classically, this can be understood in terms of an increasing number of
collision partners, making excitations in $A$ more likely to occur within
a $B$ oscillation. This manifests itself in the effective mean-field coupling
strength $g_{AB} N_A$. 
For a quantum-mechanical explanation, we firstly extract
the energy transfer time-scale $T_{\rm cycle}$ by applying compressed sensing
\cite{BergFriedlander:2008,spgl1:2007} to the $\langle \hat H_B\rangle_t$ data, 
which
gives an accurate, sparse frequency spectrum despite of the short signal length.
In fig. \ref{fig:energies} (b), we depict 
$T_{\rm cycle}=2\pi/\omega_1^{\rm CS}$ 
with $\omega_1^{\rm CS}$ denoting the angular frequency of the dominant 
peak aside from the DC peak at zero frequency. For $N_A=10$ however, the
compressed sensing spectrum is relatively smeared out such that $\omega_1^{\rm
CS}$ cannot reliably be determined. This effect is probably caused by
dephasing of various populated many-body eigenstates of $\hat H$ and so we
estimate $T_{\rm cycle}$ by the first instant $T_{\rm cycle}^{\rm max}>0$ when
$\bar \Delta^B(t)$ attains a significant local maximum.
We observe excellent agreement of $T_{\rm cycle}$ with 
the time-scale
$2\pi/\Delta E_1$
induced by the energy gap $\Delta E_1$ between the first and second excited 
many-body eigenstate of $\hat H$
obtained with
ML-MCTDHB (cf. \ref{app:method}). In order to obtain a pictorial understanding of the
$\Delta E_1$ increase with $N_A$, we have repeated our analysis for the case
$g_A=0$, allowing for the analytical expression
$\Delta E_1^{\rm pert}=
g_{AB}[(N_A-1)^2(v_{1010}-v_{0000})^2-4N_Av_{1001}^2]^{\frac{1}{2}}$
for the level spacing $\Delta E_1$
within first-order degenerate perturbation theory w.r.t. $\hat H_{AB}$. This
perturbative result is
in qualitative agreement with the numerics in fig. \ref{fig:energies} (c).
Denoting
$|n_0,n_1,...\rangle^{A}_{\rm HO}$ as a
bosonic number
state with $n_i$ atoms of type $A$ occupying $|u_i^A\rangle$,
the following mechanism becomes immediately clear: 
While for the unperturbed state $|N_A,0\rangle^{A}_{\rm HO}\otimes
|u^B_1\rangle$ a fraction of the $B$ atom wave function leaks 
more strongly into the ensemble
of $N_A$ bosons, for the unperturbed state $|N_A-1,1\rangle^{A}_{\rm HO}\otimes |u^B_0\rangle$ 
only a single $A$ boson more significantly leaks into the 
region where the $B$ is located.
Thus, the inter-species interaction raises the former state 
more in energy than the latter and thereby splits the degeneracy of these
states. Since we consider small bosonic ensembles and a weak intra-species interaction
strength, this explanation for the energy transfer acceleration
with $N_A$ holds also for $g_A=g_{AB}$ in good approximation.

Moreover, the results shown in fig. \ref{fig:energies} (a) for $N_A > 2$
feature less pronounced energy minima and maxima. 
In our terminology, the energy donation becomes less efficient and the whole
transfer cycle less complete as $N_A$ is increased.
For $N_A\geq 7$, this 
culminates in the following behaviour:
After the initial time period
of energy donation from $B$ to $A$ lasting a few tens of
$B$ oscillations, $\bar{\Delta}^B(t)$ fluctuates around the value for a balanced
distribution,
i.e. $\sim 0.6...0.65$.
We note that the above mentioned time periods
of decelerated energy transfer are not
observed for $N_A > 4$.
Instead, additional structure emerges in the form of not only deceleration but a
short time period in which $B$ accepts energy,
see e.g. the additional local minimum and maximum for
$N_A = 7$ at $t \approx 70$ and $t \approx 100$, respectively.

\section{State analysis of the subsystems}\label{sec:state_analysis}

In this chapter, we aim at a pictorial understanding of the subsystem
dynamics. For this purpose, we study the impact of the 
spatio-temporarily localized coupling on the short time-evolution of the $B$
atom density and the reduced one-body density of the $A$ species (sect.
\ref{ssec:density}). Afterwards, we employ the Husimi phase-space representation
of the
state of $B$ for investigating to which extent its initial coherence is affected
by the environment (sect. \ref{ssec:coherence}).
Likewise, we shall also characterize the environment whilst 
inspecting its depletion from a perfectly condensed
state (sect. \ref{ssec:depl_A_spec}). Further insights into the environmental
dynamics are
given
in sect. \ref{ssec:intra_spec_exc} focusing on intra-species
excitations.

\subsection{One-body densities}\label{ssec:density}

Any predictions about measurements upon the single $B$ atom alone can be
inferred from its reduced density operator $\hat \rho^B_t$, which is thus
regarded as the state of the $B$ atom:
\begin{equation}\label{eq:red_dmat_B_def}
 \hat \rho^B_t=\tr_A |\Psi_t\rangle\!\langle\Psi_t|.
\end{equation}
Here, $\tr_A$ denotes a partial trace over all $A$ bosons. Analogously, we
define $\hat \rho^A_t$ as a partial trace over
the $B$ atom and all $A$ bosons but one.

\begin{figure}[t!]
\centering
\includegraphics[width=0.75\textwidth]{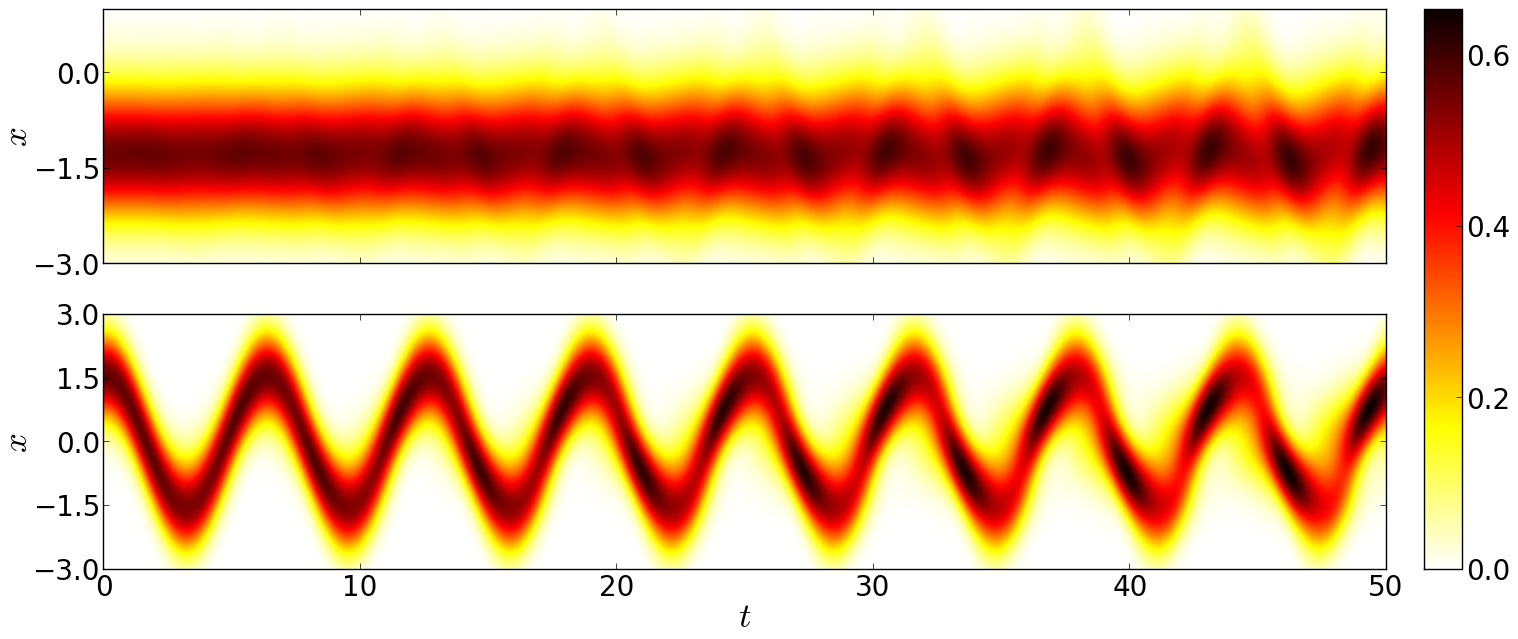}
\caption[Caption_4]{One-body densities $\rho^\sigma_t (x)$ for
the $\sigma=A$ ($B$) species in the upper (lower) panel during the first eight
B oscillations. System parameters: $N_A=2$, $\alpha=1$, $d=1.5$. All
quantities in HO units of $\hat H_B$.}
\label{fig:densities}
\end{figure}

Due to inter-species interactions, the single atom leaves a trace in the spatial
density profile of $A$.
Fig. \ref{fig:densities} shows the reduced densities, $\rho^\sigma_t (x) =
\braket{x|\hat \rho^\sigma_t|x}$ with $\sigma = A, B$,
for the first eight oscillations of $B$.
It mediates a rather intuitive picture of the process that takes place: $B$
initiates
oscillatory density modulations in $A$ via two-particle collisions emerging
already after a few $B$ oscillations.
The single atom experiences a back action from $A$ in terms of 
oscillations between
spatio-temporally localized and smeared out $\rho_t^B(x)$
density patterns.

\subsection{Coherence analysis}\label{ssec:coherence}

However, the spatial density yields no information about how coherent the state
$\hat{\rho}^B_t$ of the single atom remains. Instead, the Husimi
distribution
\begin{equation}\label{eq:husimi}
 Q^B_t(z, z^{*}) = \frac{1}{\pi} \braket{z | \hat \rho^B_t | z}, \quad z \in
\mathbb{C},
\end{equation}
serves as a natural quantity to unravel deviations from a coherent state
characterized by an isotropic Gaussian distribution.
Due to its positive-semi-definiteness, $Q^B_t(z,z^{*})$ allows for an
interpretation as the probability density to find the system in the coherent
state $\ket{z}$.
The real and imaginary parts of $z \equiv r e^{i\varphi}$ can
be identified with position and momentum in a phase space representation,
i.e. $\Re(z) = \tilde{x}/\sqrt{2}, \ \Im(z) = \tilde{p}/\sqrt{2}$. 

\begin{figure}[t!]
\centering
\includegraphics[width=0.9\textwidth]{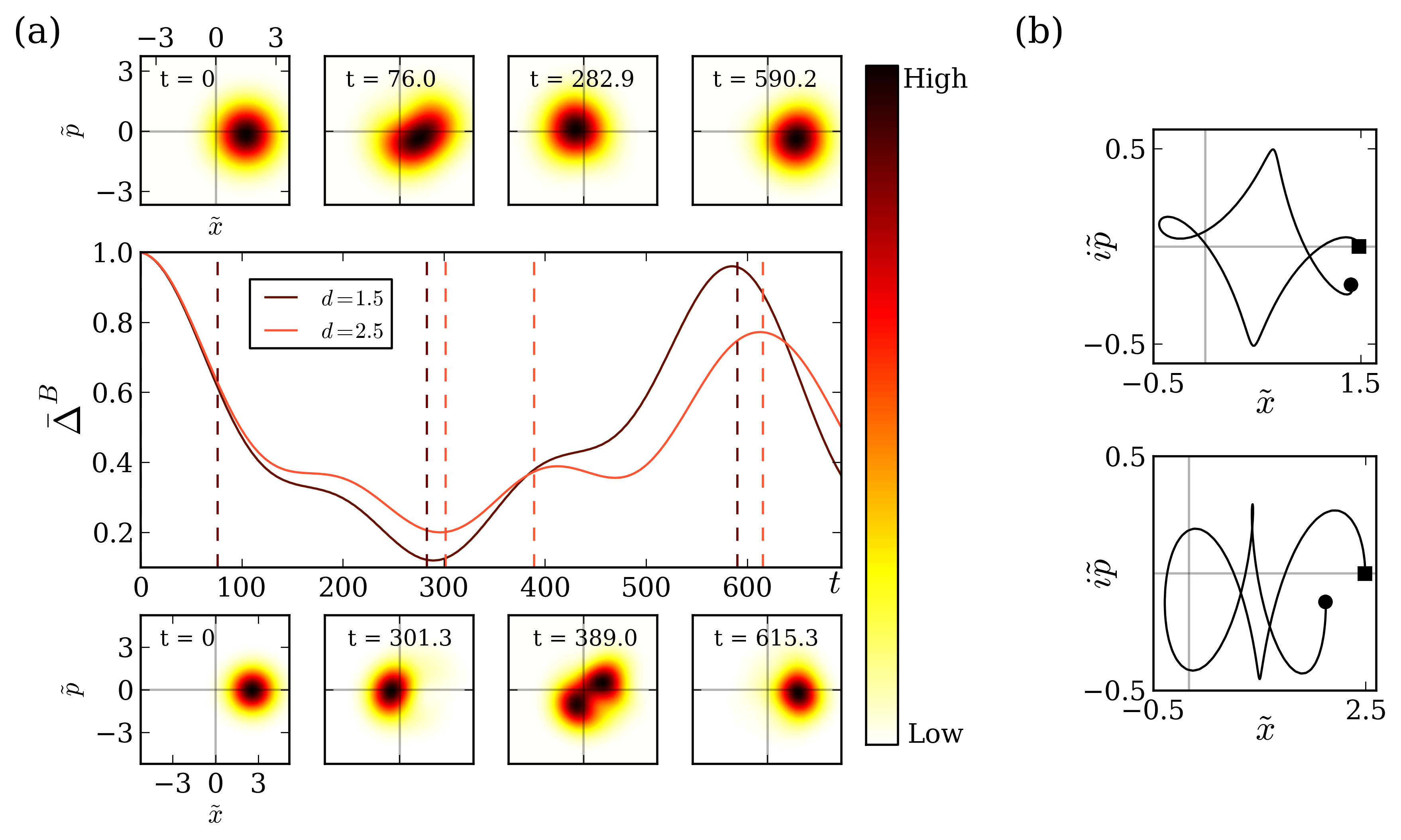}
\caption[Caption_5]{(a) Normalized excess energy $\bar \Delta^B(t)$ for $N_A =
2$ and
$d = 1.5, 2.5$.
At characteristic points in time, $\tilde{Q}^B_t(z,z^{*})$ is shown
(upper row: $d=1.5$, lower row: $d=2.5$). (b) 
$\bar z_t/\sqrt{2}$
trajectory for $d=1.5$ ($d=2.5$) until $t=590.2$ ($t=615.3$) in upper (lower)
panel. The initial (final) value of $\bar z_t/\sqrt{2}$ is indicated by a
square (circle).
All quantities
in HO units of $\hat H_B$.}
\label{fig:hus_energy}
\end{figure}

Since $B$ performs harmonic oscillations in its trap, which entail rotations of
$Q^B_t$ in phase space, the subsequent phase
space analysis is performed in the co-rotating frame of $B$ in terms of
$\tilde{Q}^B_t(z,z^{*}) \equiv Q^B_t(z e^{-it},z^{*} e^{it})$.
In fig. \ref{fig:hus_energy} (a), $\bar{\Delta}^B(t)$ is shown as well as
snapshots of
$\tilde{Q}^B_t(z,z^{*})$
at characteristic points in time for $N_A=2$ and $d = 1.5, 2.5$. The initial
Husimi
distributions are Gaussians centred around $\tilde{x} = d$.

As a matter of fact, $\tilde{Q}^B_t(z,z^{*})$ reflects the dissipative dynamics
of the $B$ atom.
The distance $\bar{r}_t$ of the mean value $\bar{z}_t$ of
$\overline{\tilde{Q}^{B}_t}(z,z^{*})$
from the origin decreases (increases) with decreasing (increasing) energy of the
$B$ atom (cf. fig. \ref{fig:hus_energy} (b)).

On the other hand, the shape of $\tilde{Q}^B_t(z,z^{*})$ provides us with
information about the quantum state of $B$.
In all our investigations for low excess energies, $\tilde{Q}^B_t(z,z^{*})$
resembles a Gaussian during most of the dynamics,
undergoing a breathing into the $\varphi$- and $r$-directions with a relatively
constant mean value $\bar z$
on the time-scale of a $B$ oscillation.
Drastic shape changes are only observed during short time periods
lasting a few $B$
oscillations 
and are accompanied with drifts of the phase $\bar\varphi_t$ of $\bar z_t$.
During these time periods, $\tilde{Q}^B_t(z,z^{*})$ features a less symmetric
shape, e.g. of a squeezed state as
for $d = 1.5$ at $t = 76.0$.
As can be inferred from fig. \ref{fig:hus_energy} (a), this short
time period coincides with a rapid energy flux from $B$ to $A$.
This suggests that the directed inter-species energy transfer is mediated
through non-coherent $B$ states.

In our frame of reference, the time-dependence of $\bar{\varphi}_t$ is due to
the
collisional coupling to the environment.
From the Husimi distribution and the $\bar z_t$ trajectory in fig.
\ref{fig:hus_energy} (b), 
we infer that $\bar{z}$ accumulates a collisional phase shift.
At $t = 282.9$, i.e. when $\bar{\Delta}^B(t)$ is minimal, a phase shift of
$\bar{\varphi}_t \approx \pi$ is observed for $d = 1.5$.
Remarkably, at $t = 590.2$, the phase shift $\bar{\varphi}_t \approx 2\pi$
indicates
that the initial state is almost fully recovered.

In the case of a displacement $d=2.5$, the excess energy $\varepsilon$ is almost three
times larger than for $d = 1.5$.
In this case, our results are of lower
accuracy (cf. \ref{app:init_converg}).
First of all, we note that the extrema of $\bar{\Delta}^B(t)$ are less
pronounced
than for $d = 1.5$ such that the recovery of the initial energy
is less incomplete.
Ultimately, one finding is similar for the higher displacement: most of the
time, $\tilde{Q}^B_t(z,z^{*})$ roughly resembles a Gaussian. Again, only
within a short
time period the shape changes drastically.
$\tilde{Q}^B_t(z,z^{*})$ then differs significantly from a Gaussian, as observed
for $d = 2.5$ at $t = 389.0$.

\begin{figure}[t!]
\centering
\includegraphics[width=0.65\textwidth]{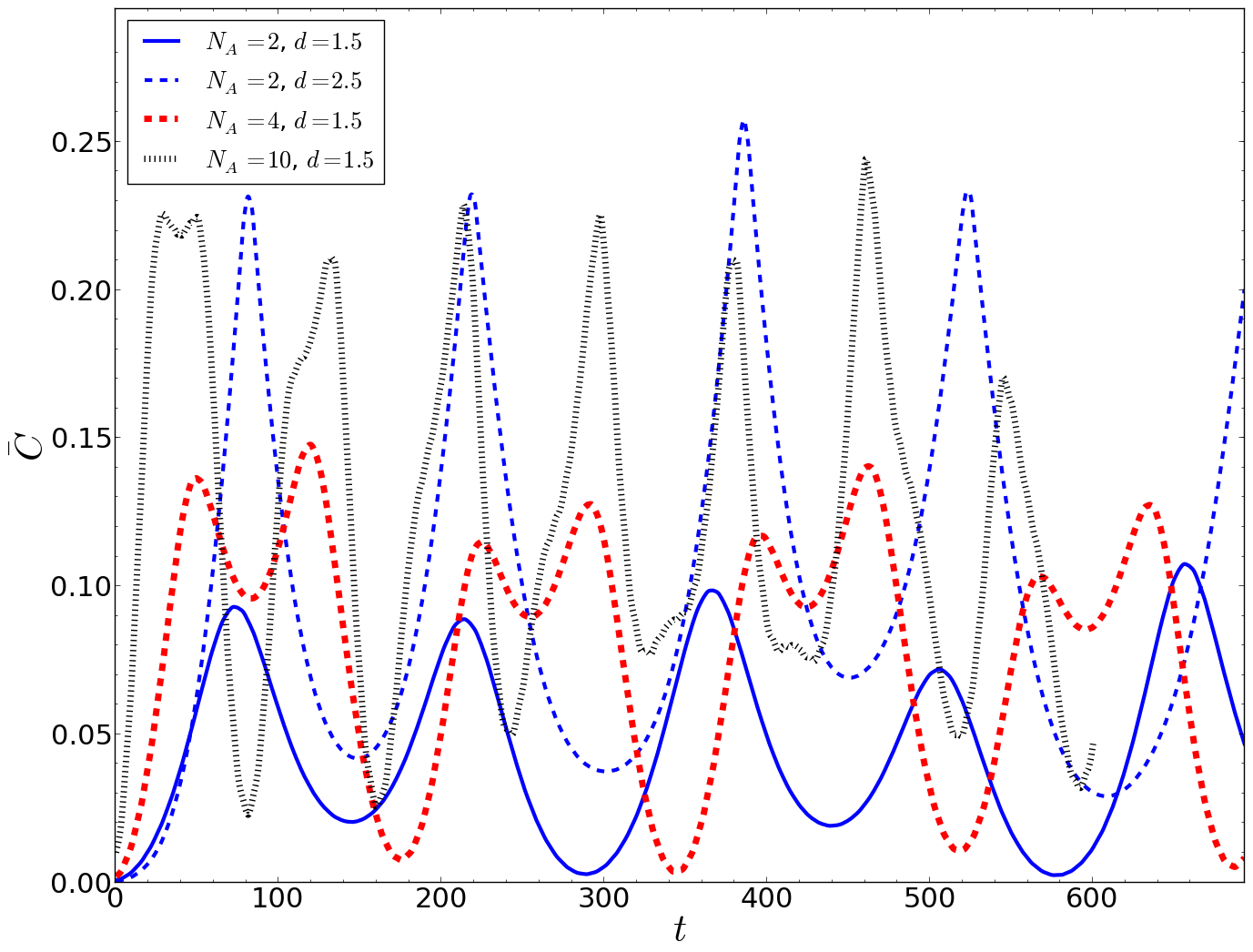}
\caption[Caption_5]{The coherence measure $\bar{C}(t)$ for 
$N_A = 2$ (blue, solid line), $N_A=4$ (red, dashed) and
$N_A=10$ (black, dotted) with $d = 1.5$. Blue dashed line: $N_A = 2$
with $d=2.5$. 
 All quantities shown in HO units of $\hat H_B$.}
\label{fig:coh_measure}
\end{figure}

In order to quantify the coherence of the $B$ quantum state (see fig.
\ref{fig:coh_measure}), we employ an
operator norm to measure the distance of $\hat{\rho}^B_t$ from its closest
coherent state:
\begin{equation}\label{eq:coh}
C_t = \frac{1}{2} \inf_z \tr ( [\hat{\rho}^B_t - |z\rangle\!\langle z|]^2
) = \frac{1}{2} \Big[ 1 + \tr\big( [\hat{\rho}^B_t]^2\big) - 2 \pi \sup_z
{Q}^B_t(z,z^{*}) \Big].
\end{equation}
We note that $C_t\in[0,1)$ and $C_t=0$ if and only if $B$ is in a coherent
state. Focusing firstly on $N_A=2$ and $d=1.5$,
the implications drawn from the
evolution of $\tilde{Q}^B_t(z,z^{*})$ persist as depicted in fig.
\ref{fig:hus_energy}:
At instants of extremely imbalanced energy distribution among $A$ and $B$, $B$
is very close to
a coherent state as indicated by $\bar{C}(t) \approx 0$ at $t \approx 289$
and $t \approx 584$.
For $d = 1.5$ and $N_A = 2$, the local minima of
$\bar{C}(t)$, e.g.
at $t \approx 145$, approximately coincide with the periods 
of decelerated energy transfer.
For larger $d$ and $N_A = 2$, the initial coherence 
is less restored 
at instants of extremal excess energy imbalance and, in between, the state of $B$
deviates more strongly from a coherent
one.
The recovery of the coherence also
becomes less complete and the maximum of $\bar C(t)$ increases 
as the size of the environment is increased while $d=1.5$ is kept fixed
(cf. $N_A = 4, 10$).
Moreover, as we will see in sect. \ref{ssec:inter_spec_corr}, 
the coherence measure $C_t$ strongly resembles the
time evolution of the inter-species correlations. This finding suggests that
the temporal deviations from a coherent state are caused by $\hat\rho^B_t$
becoming mixed.

\subsection{State characterization of the bosonic
environment}\label{ssec:depl_A_spec}

In this section, we investigate how the initially condensed state of the
finite
bosonic environment evolves structurally, thereby finding oscillations of
the $A$ species depletion, whose amplitude is suppressed when increasing $N_A$
while keeping $\varepsilon$ fixed. 
We stress that the term ``condensed'' is not used in the
quantum-statistical sense but shall refer to situations when all $N_A$ bosons
approximately occupy the same single-particle state.
For this analysis, we employ the concept of
natural orbitals (NOs) and natural populations (NPs) \cite{loewdin_norb55}, 
which are defined as the
eigenvectors and eigenvalues of the reduced density operator of a certain
subsystem in general. The spectral decomposition of the
reduced
one-body density operators for the species, $\sigma=A,B$, in particular reads:
\begin{equation}\label{eq:spectral_rho1b}
 \hat\rho^\sigma_t=\sum_{i=1}^{m_\sigma}\lambda^\sigma_i(t)\,
|\varphi^\sigma_i(t)\rangle\!\langle\varphi^\sigma_i(t)|,
\end{equation}
where $m_\sigma$ denotes the number of considered time-dependent single-particle
basis functions in the ML-MCTDHB method (cf. \ref{app:method}). Due to our
normalization of reduced density operators,
the $\lambda^\sigma_{i}(t)\in[0,1]$ add up to unity.
 In the following, we label the NPs in a decreasing sequence
$\lambda^\sigma_{i}(t)\geq\lambda^\sigma_{i+1}(t)$ if not stated otherwise. 

For characterizing the state of the bosonic
ensemble, we use the NP distribution
corresponding to the density operator of a single $A$
boson $\hat \rho^A_t$: If $\lambda_1^A(t)\approx 1$, the bosonic ensemble is
called condensed \cite{Onsager_Penrose_BEC_liquid_He_PR_1956}, whereas 
slight deviations from this case indicate
quantum depletion. Since it is conceptually 
very difficult to relate the NP distribution $\lambda_i^A(t)$ to
intra-species (and also inter-species) correlations, we may only
regard the $\lambda_i^A(t)$ as a measure for how mixed the state of a
single $A$ atom is.

\begin{figure}[t!]
\centering
\includegraphics[width=0.6\textwidth]{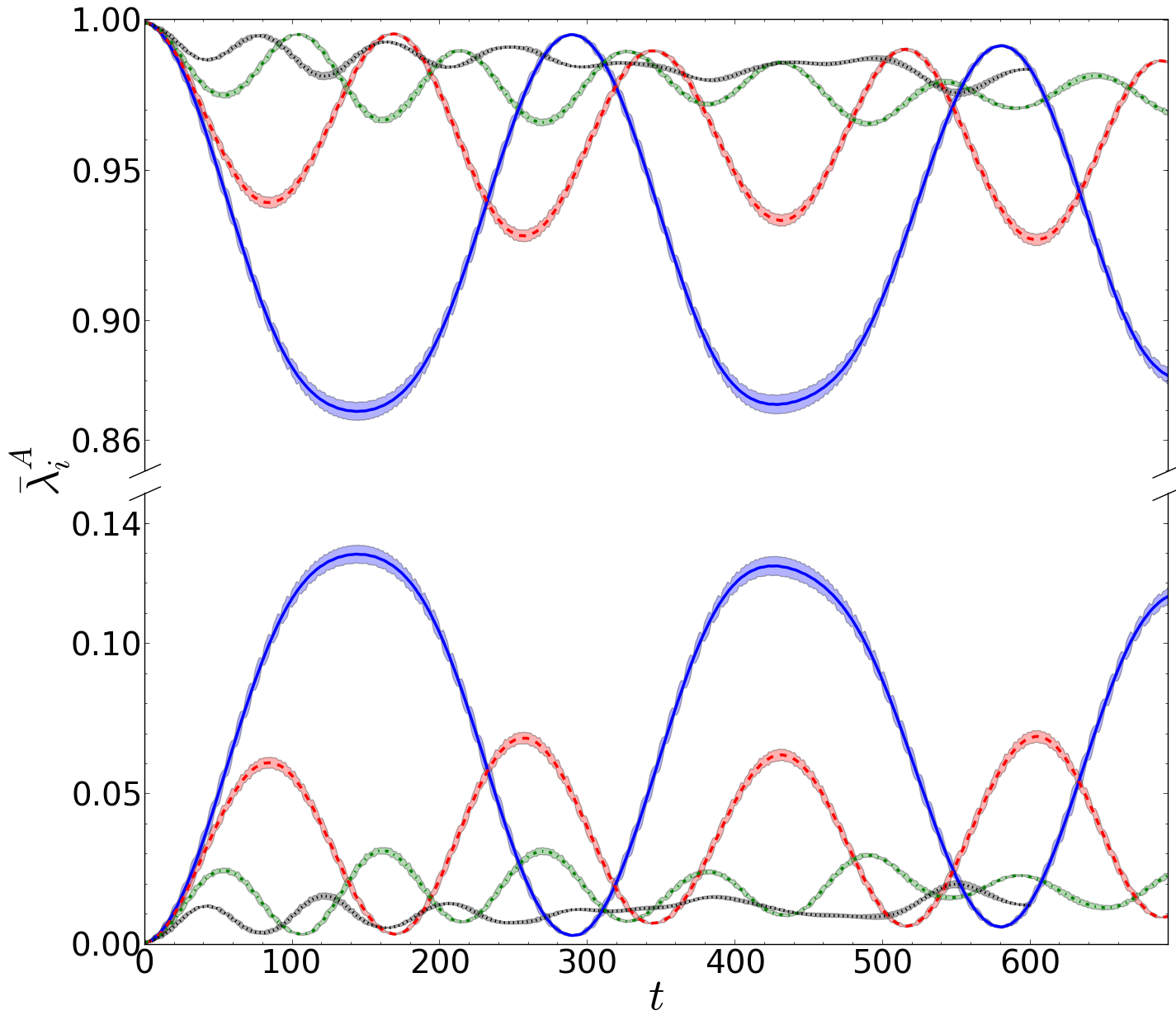}
\caption[Caption_1]{Dominant and second dominant NP of the reduced one-body
density operator of a single $A$ boson for $N_A=2$ (blue, solid line), $N_A=4$
(red, dashed),
$N_A=7$ (green, dashed dotted) and $N_A=10$ (black, dotted). 
The shaded areas indicate the standard deviation corresponding to the
$\lambda^A_i(t)$
short-time dynamics, cf.  (\ref{eq:forward_averaging_var}).
Other parameters:
$\alpha=1$, $d=1.5$. All quantities
shown in HO units of $\hat H_B$.}
\label{fig:npop_part}
\end{figure} 

For $d=1.5$, only two
NOs are
actually contributing to $\hat \rho^A_t$ - the other NPs are smaller
than $8.7\cdot10^{-3}$. Therefore, we depict only the first two NPs for
$N_A=2,4,7,10$ in fig. \ref{fig:npop_part}. Due to the weak intra-species
interaction strength, the initial depletion of the bosonic ensembles is
negligibly small, $1-\lambda_1^A(0)<10^{-3}$. One
clearly observes that the $A$ species becomes dynamically depleted and
afterwards approximately condenses again in an oscillatory manner. The
instants
of minimal depletion coincide with the instants of maximal excess energy
imbalance between the subsystems for $N_A=2$. Therefore, the two bosons behave
collectively in the sense that both approximately occupy the same
single-particle state not only during time periods 
when the $A$ species is effectively
in the ground state of $\hat H_A$ but also when $1-\bar\Delta^B(t)$ becomes maximal. 
For larger $N_A$, the depletion minima turn out to be less 
strictly synchronized with the $\bar\Delta^B(t)$ extrema. We will encounter
the same finding for the strength of
inter-species correlations in sect.
\ref{ssec:inter_spec_corr} and discuss the details there.

Strikingly, the maximal depletion is
significantly decreasing with increasing $N_A$. In \ref{app:proof_condensation},
we show that when neglecting the intra-species
interaction the higher order NPs are bounded by $\lambda^A_i(t)\lesssim
d^2/(2N_A)$, $i\geq2$, for sufficiently large $N_A$, which is a consequence
of the gapped excitation spectrum, the extensitivity of the energy of the
$A$ species and the fixed excess energy $d^2/2$. Although this
line of argument neglects intra-species interactions, which will become
important for larger $N_A$\footnote{The situation is rather involved
for bosons in a one-dimensional harmonic trap, as the ratio of $g_A$ and the
(local) linear density, which depends on $g_A$ and $N_A$, determines the
effective interaction strength.}, the numerically obtained depletions
lie well below the above bound.

\section{Correlation analysis of the subsystems}\label{sec:corr_subsystems}

As we consider a bipartite
splitting of our total
system, a Schmidt decomposition
$|\Psi_t\rangle=\sum_i\sqrt{\lambda^B_i(t)}|\Phi^A_i(t)\rangle
|\varphi^B_i(t)\rangle$ shows that the eigenvalue
distributions of $\hat\rho^B_t$ and the reduced density operator for the $A$
species,
\begin{equation}\label{eq:spectral_etaA}
 \hat\eta^A_t=\tr_B
|\Psi_t\rangle\!\langle\Psi_t|=\sum_{i=1}^{m_B}\lambda^B_i(t)\,
|\Phi^A_i(t)\rangle\!\langle\Phi^A_i(t)|,
\end{equation}
coincide. Here, we study both the emergence
of inter-species correlations on a short time-scale and their long-time
dynamics in sect. \ref{ssec:inter_spec_corr}. In particular, we show 
analytically that these correlations 
essentially vanish whenever the excess energy
$\varepsilon^\sigma_t$ of one species $\sigma=A,B$
is much lower than the
excitation gap of $\hat H_\sigma$. In sect. \ref{ssec:E_NORBs}, 
we then unravel the energy transfer 
between the species
by inspecting the energy stored in the respective NOs $|\Phi^A_i(t)\rangle$, 
$|\varphi^B_i(t)\rangle$ of the subsystems. Thereby, we identify 
incoherent energy
transfer processes, which are related to inter-species correlations and 
shown to accelerate the early energy donation of the $B$ atom to the $A$
species.

\subsection{Inter-species correlations}\label{ssec:inter_spec_corr}

The NPs $\lambda_i^B(t)$ are directly connected to
inter-subsystem correlations: Since the $B$ species consists of a single atom 
being distinguishable from the $A$ atoms, and since the
bipartite system always stays in a pure state, the initial pure state of the
$B$ atom characterized by $\lambda_1^B(0)=1$ can only become mixed if
inter-species correlations are present. Deviations from $\lambda_1^B(t)=1$
indicate entanglement between the single atom and the species of $A$ bosons. 
In order to quantify these inter-subsystem correlations,
we employ the von Neumann entanglement entropy:
\begin{equation}\label{eq:vN_entropy_def}
S_{\rm vN}(t) = -{\rm tr}\left( \hat\rho^B_t\,\log \hat\rho^B_t\right)  
	      = - \sum_{i=1}^{m_B}\lambda_i^B(t)\,\log \lambda_i^B(t),
\end{equation}
which vanishes if and only if inter-species correlations are absent, i.e.
$|\Psi_t\rangle=|\Phi^A_1(t)\rangle|\varphi^B_1(t)\rangle$.

\begin{figure}[t!]
\centering
\includegraphics[width=0.9\textwidth]{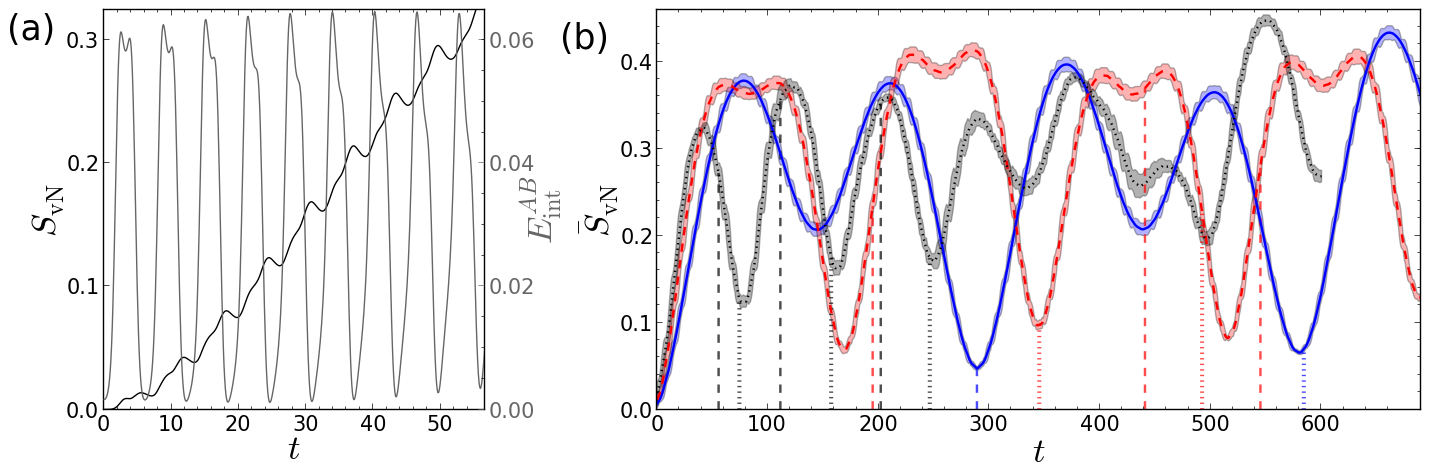}
\caption[Caption_1]{(a) Short-time evolution of the
von Neumann entanglement entropy $S_{\rm vN}(t)$ (black line, left
ordinate) and 
inter-species
interaction energy $E^{AB}_{\rm int}(t)=\langle \hat H_{AB}\rangle_t$ 
(grey, right ordinate)
for $N_A=2$ (data not time-averaged). (b) Long-time evolution of 
$\bar S_{\rm vN}(t)$
for $N_A=2$ (blue solid line), $N_A=4$ (red, dashed) and $N_A=10$ (black,
dotted). 
The vertical dashed (dotted) lines in the respective colour indicate the
instants of local $\bar\Delta^B(t)$ minima (maxima) read off fig.
\ref{fig:energies} (for clarity, we only show these reference lines until
$t=320$
for $N_A=10$).
The shaded areas indicate the standard deviation of $S_{\rm vN}(t)$
from $\bar S_{\rm vN}(t)$.
Other parameters for both subplots: $\alpha=1$, $d=1.5$. All quantities
shown in HO units of $\hat H_B$.}
\label{fig:entropy}
\end{figure}

For all considered $N_A$, the von Neumann entropy of both the ground 
state of $\hat H$ and our initial condition
$|\Psi_0\rangle$ with $d=1.5$ does not
exceed $3.5\cdot10^{-3}$, 
which is negligible compared to the dynamically attained values depicted
in fig. \ref{fig:entropy}. Thus, we may conclude that (i) the
spatio-temporarily localized
coupling requires a certain amount of excess energy for the subsystems to
entangle and (ii) inter-species correlations are dynamically established
via interferences involving excited states.

In fig. \ref{fig:entropy} (a), we present the short-time 
dynamics of the $S_{\rm vN}(t)$
for $N_A=2$. The entanglement between the single $B$ atom
and the two $A$ bosons is built up in a step-wise manner with each collision.
Since the local maxima of $S_{\rm vN}(t)$ are delayed w.r.t. the
corresponding maxima of $\langle \hat H_{AB}\rangle_t$, we conclude that a finite
interaction time is required in order to enhance correlations. 
After a few $B$ oscillations, the effective interaction time is
increased since (i) dissipation moves the left turning point of the $B$ atom
from $x=-d<-R$ towards the centre of the $A$ species at $x=-R$ and (ii) the
$A$ / $B$ species density oscillations become synchronized (cf. fig.
\ref{fig:densities}). As a consequence, the overall slope of $S_{\rm vN}(t)$
increases during the first few $B$ oscillations. We note that such a step-wise
emergence of correlations has been reported also in
\cite{mack_dynamics_2002} for two atoms colliding in a single harmonic trap.

Concerning the long-time dynamics, we first focus on the $N_A=2$ case in fig.
\ref{fig:entropy} (b). The correlation increase continues until a maximum at
$t\approx78$, which is followed by a minimum around $t\approx 144$ of 
modest depth, a further maximum at $t\approx210$ and a deep minimum at 
$t\approx288$. This alternating sequence of maxima and minima is 
repeated thereafter. In passing, we
note that to some extent similar entropy oscillations with a long period
comprising many collisions
have been reported for two atoms in a single harmonic trap interacting with
repulsive and attractive contact interaction in \cite{mack_dynamics_2002} and
\cite{sowinski_dynamics_2010}, respectively.
As indicated by the vertical lines, the deep minima of 
$\bar S_{\rm vN}(t)$ at $t\approx288$ and $t\approx580$ coincide very well with
instants of both a maximally imbalanced excess energy distribution among
the species (cf. fig. \ref{fig:energies}) and minimal
$A$ species depletion (cf. fig. \ref{fig:npop_part}).

This intimate relationship between maximal excess energy imbalance and absence
of 
correlations can be understood analytically by expressing 
$\langle \hat H_B\rangle_t$ as a function(al) of the NPs and NOs,
\begin{equation}\label{eq:EB_NORB_unravelling}
 \langle \hat H_B\rangle_t = \sum_{i=1}^{m_B}
\lambda_i^B(t)\,\langle\varphi^B_i(t)|\hat H_B|\varphi^B_i(t)\rangle,
\end{equation}
and identifying $\Sigma^B_i(t)=\langle\varphi^B_i(t)|\hat H_B|\varphi^B_i(t)\rangle$
with the energy content of the $i$-th NO.
In \ref{app:proof_disentanglement}, we
show that $\varepsilon^B_t$ being much smaller than the excitation 
gap $(E^B_1-E^B_0)$ implies $\lambda_i^B(t)\ll 1$ for $i>1$. At those instants, 
the presence of
an excitation gap prevents correlations between the subsystems due
to a lack of orthonormal states $|\varphi^B_i(t)\rangle$, $i>1$, 
with excess energy $\varepsilon^B_i(t)=\Sigma^B_i(t)-E^B_0$ being of 
$\mathcal{O}(\varepsilon^B_t)$. The same line of arguments analogously holds 
also for
instants when $\varepsilon^A_t$ is much smaller than the excitation gap of 
$\hat H_A$. In between instants of maximal excess energy imbalance, more states are
energetically accessible such that inter-species correlations can be
established. Increasing $N_A$ while keeping $d=1.5$ alters the $\bar S_{\rm
vN}(t)$ dynamics in the following way:

(i) Inter-species correlations initially emerge faster 
because binary collisions become more likely within a $B$ oscillation (cf.
$N_A=2,4,10$ curves for $t<35$). 

(ii) The depth of shallower 
minima\footnote{For $N_A=1$, these minima even attain a depth
similar to the deep minima (plot not shown).} 
decreases (cf. the $N_A=4$ curve at e.g.
$t\approx83$) such that neighbouring pairs of local maxima (cf. the $N_A=4$ 
curve at e.g.
$t\approx60$ and $t\approx107$) degenerate to a single maximum for $N_A>4$ 
(all entropy minima for e.g. $N_A=10$ correspond to the deep minima
observed for smaller $N_A$).

(iii) We find the coincidence of the deep entropy minima with extrema
in the excess energy imbalance $\bar\Delta^B(t)$ to be less\footnote{
We note that the deep  $\bar S_{\rm vN}(t)$  minima remain synchronized
with
the $A$ species depletion minima for all considered $N_A$ (for $N_A=10$ and
$t>300$ the depletion minima, however, are
quite washed out).}
 established. For e.g. $N_A=4$, the first
deep
$\bar S_{\rm vN}(t)$ minimum at $t\approx169$ is attained before the 
first $\bar\Delta^B(t)$ minimum at $t\approx195$.
Since the overall energy donation of the $B$ atom is less efficient
for $N_A=4$ (compared to $N_A=2$), $\bar S_{\rm vN}(t)$ is not forced to
attain a deep minimum at the first minimum of $\bar\Delta^B(t)$ in the sense of
the
above analytical line of argument. For $N_A=10$, 
$\bar\Delta^B(t)$ essentially fluctuates around $0.65$ for $t>60$ such that
the excitation gaps do not impose any restrictions on $\bar S_{\rm vN}(t)$.

(iv) The minimal value of  $\bar S_{\rm vN}(t)$ for $t>0$ increases, which goes 
hand in hand with the energy donation of the $B$ 
atom becoming less efficient and the energy transfer cycle becoming less
complete.

(v) The maximal values of $\bar S_{\rm vN}(t)$ are all comparable for
$N_A\leq10$.
We note that the persistence of inter-species correlations when increasing
$N_A$ does not contradict the fact that the $A$ species becomes simultaneously
more
condensed (cf. sect. \ref{ssec:depl_A_spec}), which can be
easily seen by a minimal example provided in \ref{app:example_persist_corr}.

\subsection{Energy distribution among natural
orbitals and incoherent transfer processes}\label{ssec:E_NORBs}

In this section, we unravel the impact of the previously identified
inter-species correlations on the energy transfer and identify incoherent
transfer processes. For this purpose,
we will evaluate the contribution of $\Sigma^B_i(t)$ to $\langle
\hat H_B\rangle_t$ and complement this energy decomposition by analysing how the
energy of the $A$ species is distributed among the $A$ species NOs
$|\Phi^A_i(t)\rangle$:
\begin{equation}\label{eq:Aspec_energy_distr}
 \langle \hat H_{A}\rangle_t = \sum_{i=1}^{m_B}\lambda^B_i(t)\,\Sigma^A_i(t).
\end{equation}
Here, we have introduced the energy content of the $i$-th $A$ species NO:
$\Sigma^A_i(t)=\langle \Phi^A_i(t)|\hat H_A|\Phi^A_i(t)\rangle$. 
We remark that the
presence of intra-species interaction prevents us from expressing $\langle
\hat H_{A}\rangle_t$ as a function(al) of the NPs and NOs of $\hat\rho^A_t$
in a similar fashion as in (\ref{eq:EB_NORB_unravelling}). The 
highly challenging task of diagonalizing the reduced density operator $\hat
\eta^A_t$ of the whole $A$ species for obtaining the $N_A$-body states
$|\Phi^A_i(t)\rangle$, however, can be faced efficiently with the ML-MCTDHB
method due to its beneficial representation of the many-body wave function (cf. 
\ref{app:method}). We note that the NO
energy contents $\Sigma^{A/B}_i(t)$ constitute system-immanent,
basis-independent quantities as they result from the diagonalization of reduced
density operators.
\begin{figure}[t!]
\centering
\includegraphics[width=0.75\textwidth]{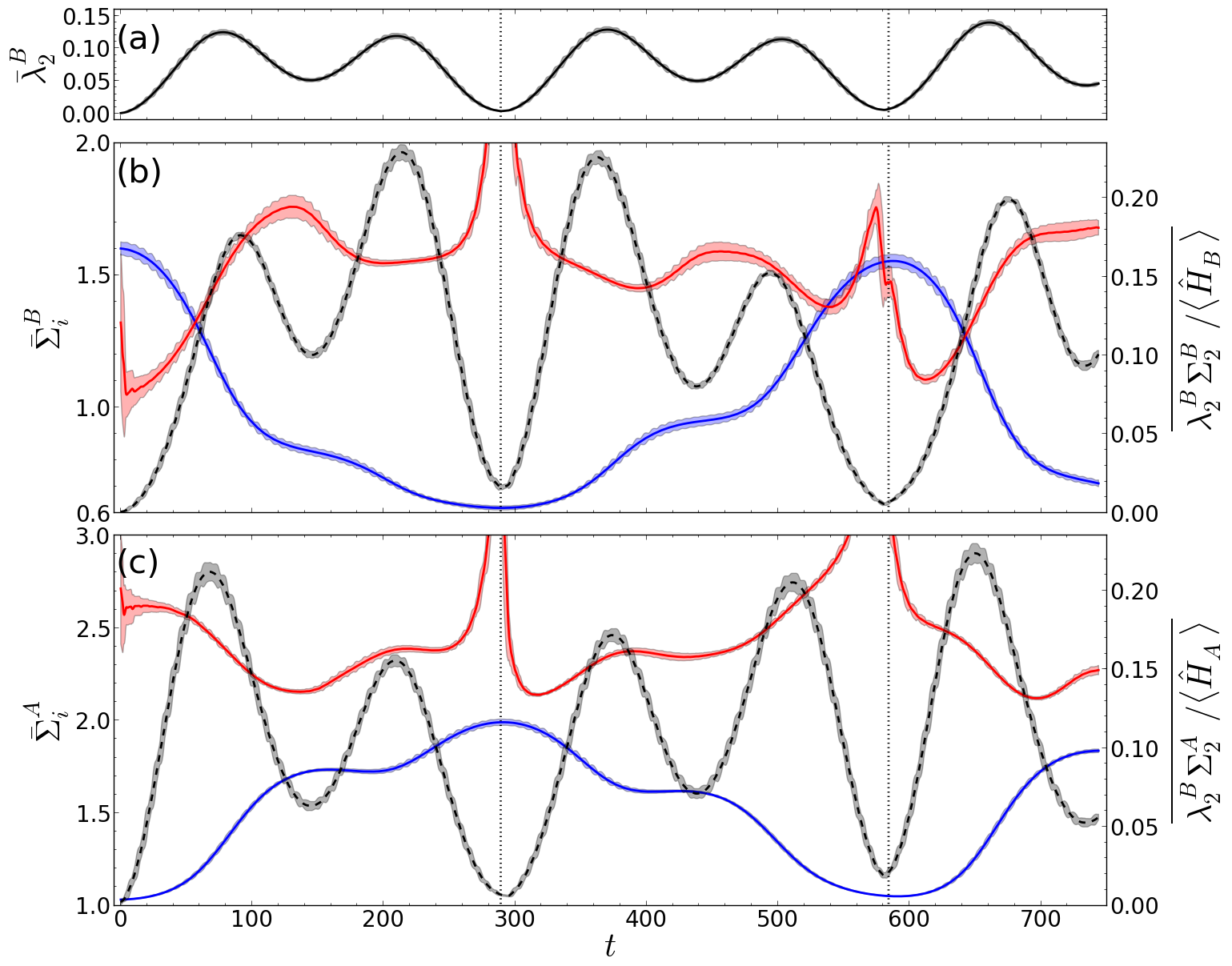}
\caption[Caption_1]{(a) Second dominant NP of
$\hat\rho^B_t$. (b)/(c) NO energy content $\bar\Sigma^\sigma_i(t)$ of
the first ($i=1$, blue solid line) and second ($i=2$, red) NO of species
$\sigma =A/B$ (left ordinate). Black dashed line: 
Weighted contribution of the second dominant NO to the respective
intra-species energy (right ordinate). The vertical dotted lines at $t\approx
289$ ($t\approx 584$) indicate
the instants of minimal (maximal) $\bar\Delta^B(t)$. 
The shaded areas indicate the standard deviation from the respective locally
time-averaged quantity.
System parameters:
$N_A=2$, $\alpha=1$, $d=1.5$. All quantities
shown in HO units of $\hat H_B$.}
\label{fig:norb_E_content}
\end{figure}

For the considered displacement $d=1.5$ and
all
considered $N_A$, we find that only two NOs contribute to $\hat\rho^B_t$
and $\hat\eta^A_t$ ($\lambda^B_i \lesssim 7.6\cdot10^{-3}$ for
$i>2$) so that we only analyse their energy contents in fig.
\ref{fig:norb_E_content}. Larger displacements generically result in more
populated NOs and thus a much more involved dynamics, whose analysis goes
beyond the scope of this work. 
In fig. \ref{fig:norb_E_content}, we
bring the evolution of the second NP $\bar\lambda^B_2(t)$ face to face with the
energy
contents of the dominant and the second dominant NOs $\bar\Sigma^{A/B}_i(t)$,
$i=1,2$, for
the case of $N_A=2$ bosons. The NP of the dominant NO can be inferred from
$\bar\lambda^B_1(t)\approx 1- \bar\lambda^B_2(t)$. Since
$\bar\Sigma^{\sigma}_i(t)$ does not contain information about
how much energy of the $\sigma$ species is actually stored in the
corresponding NO, we moreover quantify the contribution of the second NO to
the subsystem energy in terms of the ratio
$\overline{\lambda^B_2\Sigma^{\sigma}_2
/\langle\hat H_{\sigma}\rangle}(t)$. 

We clearly observe that the energy content of the dominant NO,
$\bar \Sigma^{A/B}_1(t)$, qualitatively resembles the subsystem energy
$\bar H_{A/B}(t)$ evolution (cf. fig. \ref{fig:energies}). However, at 
time periods when
inter-species correlations are present, i.e. when $\bar\lambda^B_2(t)\gtrsim
3\cdot10^{-2}$, a significant fraction of the subsystem energy is stored in the
respective second dominant NO. This fraction can become as large as
$23\%$. Since essentially all
subsystem energy is stored in the respective dominant NO whenever 
$\bar\Delta^B(t)$ is extremal (cf. at $t\approx 289$ and $t\approx 584$) and
since the evolution
of $\overline{\lambda^B_2\Sigma^{A}_2
/\langle\hat H_{A}\rangle}(t)$
appears to be
synchronized with the dynamics of $\overline{\lambda^B_2\Sigma^{B}_2
/\langle\hat H_{B}\rangle}(t)$, we conclude that the 
energy transfer is mediated via incoherent
processes in the following sense: Rather than keeping all its instantaneous
energy in the dominant NO, the $B$ atom shuffles energy from the dominant to the
second NO and back while donating (accepting) energy to (from) the $A$
species. The $A$ species, in turn, accepts (donates) energy from (to) the $B$
atom while shuffling energy from the dominant to the second NO and back.
As predicted by
the analytical line of argument in \ref{app:proof_disentanglement}, one can
furthermore witness how the intra-species
excitation gap of the $\sigma$
species makes the second NO to energetically separate
drastically from the dominant one when
$\varepsilon^\sigma_t$ becomes too small (cf. $\bar\Sigma^B_2(t)$ at
$t\approx289$ and $\bar\Sigma^A_2(t)$ at
$t\approx584$). Finally, we observe that
$\bar\Sigma^{A/B}_2(t)>\bar\Sigma^{A/B}_1(t)$ holds most of the time.
Consequently, if the $B$ atom has been detected in the NO of lower (higher)
energy, the $A$ species is found almost certainly in the NO of lower (higher)
energy and vice versa, according to the above Schmidt decomposition.

The above overall picture of the NO energy content
evolution remains valid
for all $N_A\leq 10$, while the relationship between the inter-species energy
transfer and the distribution of the subsystem energies among the respective NOs
becomes more involved for larger environments. Whereas
$\overline{\lambda^B_2\Sigma^{A/B}_2
/\langle\hat H_{A}\rangle}(t)$ remains synchronized with
$\bar S_{\rm vN}(t)$ when going to larger $N_A$, the
inter-species correlation dynamics becomes less strictly synchronized with the
$\bar\Delta^B(t)$ evolution as discussed at the end of sect.
\ref{ssec:inter_spec_corr} in detail. With increasing $N_A$, we moreover
observe the tendency that each species temporarily stores a slightly more
significant fraction of its energy also in the third and fourth dominant NO
despite their quite low populations. We suspect that this increase in complexity
might be related to the observed decay of the excess energy imbalance 
to a rather balanced distribution for $N_A=7, 10$ (cf. fig.
\ref{fig:energies}). Finally, we summarize peculiarities concerning $N_A=2$, 
which are diminished or
absent for larger environment sizes: Firstly, the decelerated energy transfer
around $t\approx 160$ coincides with a time period
of significant contribution of the
second NO
to $\bar H_B(t)$ so that the incoherent processes can be made partially 
responsible for the delay. Secondly, pairs of subsequent
$\overline{\lambda^B_2\Sigma^{A/B}_2
/\langle\hat H_{A/B}\rangle}(t)$
maxima in between consecutive deep
minima (e.g. at $t\approx91$ and
$t\approx214$) are related
to the observed pairs of local $\bar S_{\rm vN}(t)$ maxima and, thus, merge to
a single maximum for larger $N_A$.

Finally, we analyse how the overall energy transfer is influenced by these 
incoherent energy transfer processes.
The ML-MCTDHB method allows to manually switch
off inter-species correlations in order to clarify their role for the dynamics
(cf. \ref{app:method}).
By comparison, we have found for all considered
$N_A$ that the presence
of inter-species correlations accelerates the energy donation
of the $B$ atom at first, i.e. for $t\lesssim50$,
which is plausible since more energy transfer channels are open. Whether
the 
overall energy donation is accelerated, however,
depends on $N_A$: For $N_A=2$, we observe an acceleration by a factor of
two, while for $N_A=7$ no overall acceleration is found (plots not shown).

\section{Excitations in Fock space and their correlations}\label{sec:exc_fock}

Operating in the weak interaction regime, we naturally
define intra-system excitations as occupations of excited single-particle
eigenstates corresponding to the respective HO
Hamiltonian. The joint probability
for finding $(n_0,n_1,...)$ $A$ bosons in the respective HO
eigenstates and the $B$ atom in its $j$-th HO eigenstate is given by:
\begin{equation}\label{eq:def_joint_fock_prob}
 P_t(n_0,n_1,...;j)=\left\vert\,\langle\Psi_t|\left(|n_0,n_1,...\rangle^{A}_{\rm
HO} \otimes|u_j^B\rangle\right)\,\right\vert^2.
\end{equation}
In this chapter, we firstly show that the actual long-time excitation dynamics
takes
place only in certain active subspaces being mutually decoupled (sect.
\ref{ssec:subspaces_fixed_SP_E}). Secondly, the $A$ species
excitations are shown to be governed by singlet and delayed doublet
excitations (sect. \ref{ssec:intra_spec_exc}). The findings of these two
sections are explained by means of a tailored time-dependent perturbation
theory. Finally, correlations between the intra-species excitations of
$A$ and $B$ are shown to (dis-)favour certain energy transport channels 
depending, in general, on the direction of energy transfer (sect.
\ref{ssec:fock_corr}).

\subsection{Decoupled active subspaces}\label{ssec:subspaces_fixed_SP_E}
For a fixed total single-particle energy $E_{\rm SP}(k)=k+(N_A+1)/2$,
$k\in\mathbb{N}_0$, we have evaluated the probability $P_{\rm SP}(k;t)$
to find the bipartite system in the subspace spanned by configurations
$|n_0,n_1,...\rangle^{A}_{\rm HO}\otimes|u^B_j\rangle$ with
total single-particle energy $E_{\rm SP}(k)$. So $P_{\rm
SP}(k;t)$ is obtained by summing over all $P_t(n_0,n_1,...;j)$ with
$j+\sum_{r\geq 1}r\,n_r=k$. In fig. \ref{fig:fock_correlations} (a), we
depict the probabilities $\bar P_t(n_0,n_1,...;j)$ for the configurations of
total single-particle energy $E_{\rm SP}(1)$, $|N_A,0,...\rangle^{A}_{\rm
HO}\otimes|u^B_1\rangle$ and $|N_A-1,1,...\rangle^{A}_{\rm
HO}\otimes|u^B_0\rangle$, for $N_A=4$, showing large amplitude oscillations
while
their
sum $\bar P_{\rm SP}(1;t)$ (not shown) stays comparatively constant. 
In fact, we find
$\bar P_{\rm SP}(k;t)$ to be approximately conserved\footnote{This
approximation becomes less valid for larger $N_A$. For $N_A=10$,
$\bar P_{\rm SP}(k;t)$ features slight drifts on a long time-scale.} for all
considered $N_A$ and $d=1.5$. Thus, the actual
energy transfer dynamics approximately happens solely 
within the various subspaces of fixed total single-particle energy $E_{\rm
SP}(k)$, 
$k\geq1$\footnote{The subspace with $k=0$
is the only one-dimensional one, i.e. $P_{\rm
SP}(k,t)\approx P_t(N_A,0,...;t=0)$.},
while inter-subspace transitions are suppressed. 

\subsection{Intra-species excitations}\label{ssec:intra_spec_exc}
The
joint probability (\ref{eq:def_joint_fock_prob}) describing the distribution of
excitations in the total
system defines the following two marginal distributions for the $A$ species and
the single $B$ atom,
\begin{eqnarray}\label{eq:marginal_A}
 p^A_t(n_0,n_1,..)&=\sum_j P_t(n_0,n_1,...;j)={}^{\;\;\;\,A}_{\;\rm HO}\langle
n_0,n_1,...|\hat \eta^{A}_t
|n_0,n_1,...\rangle^{A}_{\rm HO},\\
p^B_t(j)&=\sum_{n_0,n_1,...} P_t(n_0,n_1,...;j)=\langle u^B_j|\hat \rho^B_t
|u^B_j\rangle.
\end{eqnarray}
In order to classify the collisionally induced excitations in the
environment, we 
inspect the probabilities for having no ($p_t^{A;0}$),
a singlet ($p_t^{A;1}$) and a doublet excitation ($p_t^{A;2}$),
\begin{equation}\label{eq:no_exc_A}
 p_t^{A;0}=p^A_t(\vec n_0),\quad
\label{eq:sing_exc_A}
 p_t^{A;1}=\sum_{i>0} p^A_t(\vec n_0^i),\quad
\label{eq:doub_exc_A}
 p_t^{A;2}=\sum_{0<i\leq j} p^A_t(\vec n_0^{ij}),
\end{equation}
where $\vec n_0=(N_A,0,0,...)$ and $\vec n_0^i$ ($\vec n_0^{ij}$) refers to
occupation number vectors with $N_A-1$ ($N_A-2$) atoms in the harmonic
oscillator ground state and one (two) atom(s) in the $i$-th ($i$-th and
$j$-th) excited state(s). 
\begin{figure}[t!]
\centering
\includegraphics[width=0.6\textwidth]{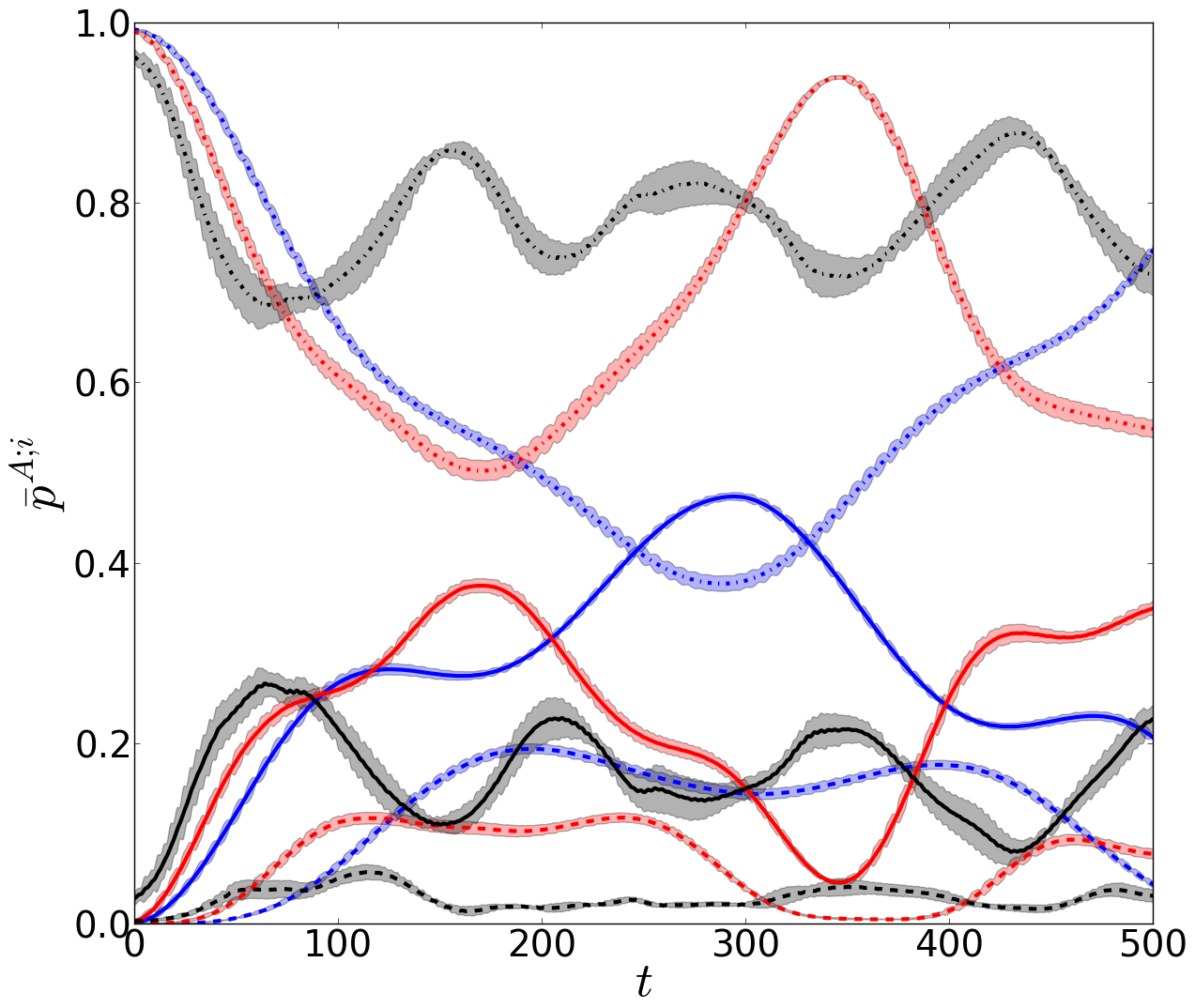}
\caption[Caption_1]{Probability $\bar p^{A;i}(t)$ for no ($i=0$, dashed dotted
line), singlet ($i=1$, solid) and
doublet ($i=2$, dashed) excitation(s) for $N_A=2$ (blue), $N_A=4$ (red) and 
$N_A=10$
(black).
The shaded areas indicate the standard deviation from the respective locally
time-averaged quantity.
Other system parameters: $\alpha=1$, $d=1.5$.
All quantities shown in HO units of $\hat H_B$.
}
\label{fig:exc_class}
\end{figure}
In fig. \ref{fig:exc_class}, we compare these classes of $A$ species
excitations
for $N_A=2,4,10$: One can clearly see that the excitations attain a maximum
(minimum)
whenever $\bar\Delta^B(t)$ (cf. fig. \ref{fig:energies}) becomes minimal
(maximal). Moreover, singlet 
excitations dominate over doublets (and also not shown higher order
excitations),
which occur only after a certain delay with respect to the appearance of
singlets. As expected from energetical considerations, the singlet excitations
involve dominantly the $|N_A-1,1,0,...\rangle^{A}_{\rm HO}$ configuration and
second dominantly $|N_A-1,0,1,0...\rangle^{A}_{\rm HO}$, which can occur
as likely as the dominant doublet contribution $|N_A-2,2,0,...\rangle^{A}_{\rm
HO}$.

In sect. \ref{sec:energy}, the increasing number of collision partners has
been argued to be responsible for the acceleration of the energy transfer with
increasing $N_A$. Fig. \ref{fig:exc_class} clearly shows this enhancement of
excitation probability with increasing $N_A$ during the first few $B$
oscillations ($t\lesssim80$). The larger $N_A$, the faster does the probability
for singlet 
excitations increase. For $N_A=10$, this
effect is less visible since the intra-species
interaction results in a finite initial singlet probability.
Concerning the long-time dynamics, we observe that the maximal singlet / doublet
excitation probability reduces for larger environment sizes, which goes hand in
hand with the energy donation to A becoming less efficient.

For obtaining analytical insights into the delayed doublet emergence and the
decoupled active subspaces, we have applied a stroboscopic version of
time-dependent perturbation theory to the simplest case of $g_A=0$. The total
wave function $|\Psi_r\rangle$ after $r$ oscillations is propagated for
one free oscillation period $T=2\pi$ with linear perturbation theory with
respect to $\hat H_{AB}^I(t)$ (with $I$ denoting the interaction picture):
$|\Psi^I_{r+1}\rangle=(\mathds{1}-i2\pi \bar{H}_{AB}^I)|\Psi^I_r\rangle$, where
the time-averaged coupling Hamiltonian reads
\begin{equation}\label{eq:strob_td_purb_hamilt}
 \bar{H}_{AB}^I = \frac{1}{2\pi}\int\limits_{2\pi r}^{2\pi (r+1)}{\rm
d}\tau\,\hat H_{AB}^I(\tau)=
g_{AB} \sum_{i,j,p,q} \delta_{i+j,q+p} \,v_{ijpq} \,{\hat a}^\dagger_i {\hat
a}_p \otimes |u^B_j\rangle\!\langle u^B_q|.
\end{equation}
The energy conservation enforced by the Kronecker $\delta_{i+j,q+p}$, 
being reminiscent of Fermi's golden rule,
is an immediate consequence of
the equidistant single-particle spectra and $\alpha=1$. In order to avoid norm
loss, we raise the above map to a unitary one,
$|\Psi^I_{r+1}\rangle\approx\exp(-i\,2\pi \bar{H}_{AB}^I)|\Psi^I_r\rangle$,
which is equivalent to solving a temporally coarse-grained Schr\"odinger
equation
with the effective Hamiltonian $\bar{H}_{AB}^I$.

We have observed good agreement between full numerical
simulations (with $g_A=0$) and the perturbation theory (plots not shown)
concerning the observables $\bar\Delta^B(t)$ and $\bar p^{A;i}(t)$ during
the first $18$ ($12$) collisions for $N_A=2$ ($N_A=7$). Yet the entanglement
entropy turns out to be overestimated a bit earlier by the perturbative
approach.
Comparing
our full numerical data for $g_A=0$ and $g_A=g_{AB}$, we find qualitative
agreement 
during the first $r=10,...,20$ collisions such that the following insights from
the perturbation theory should also apply in the presence of intra-species
interaction. Firstly, one can easily show that $\bar{H}_{AB}^I$ commutes with
the projector onto the subspace of configurations with fixed $E_{\rm SP}(k)$.
Due to the detailed energy conservation in binary collisions, the temporally
coarse-grained dynamics decouples subspaces of different total
single-particle energy. Secondly, $\bar{H}_{AB}^I$ involves only one-body
processes
within the $A$ species. Therefore, doublets can only emerge after the second
collision with - as second order processes w.r.t. to  $\bar{H}_{AB}^I$
- smaller amplitude than singlets, which explains the suppression and delay of
the doublet creation. It goes without saying that the perturbative approach
breaks down after fewer collisions for larger $N_A$ since (i) the intra-species
interaction becomes more relevant and (ii) the enhanced inter-species coupling
$g_{AB}N_A$ leads to less separated time scales making a temporal
coarse-graining questionable.

\subsection{Correlations between intra-species
excitations}\label{ssec:fock_corr}
Ultimately, we aim at unravelling correlations
between excitations of the $B$ atom and the $A$ species. For this
purpose, we introduce the following correlation measure:
\begin{equation}\label{eq:g2_fock_def}
 g_t(n_0,n_1,...;j)=\frac{P_t(n_0,n_1,...;j)}{p^A_t(n_0,n_1,..)\,p^B_t(j)},
\end{equation}
which is reminiscent of the diagonal of higher order coherence functions
introduced by Glauber \cite{glauber_quantum_1963}. While $g_t(n_0,n_1,...;j)=1$ 
means that the intra-species
excitations take place statistically independently, a value
larger (smaller) than unity implies that the corresponding joint excitation 
event $(n_0,n_1,...;j)$ is measured with an enhanced (decreased) probability,
i.e. (anti-)bunches.
\begin{figure}[t!]
\centering
\includegraphics[width=0.8\textwidth]{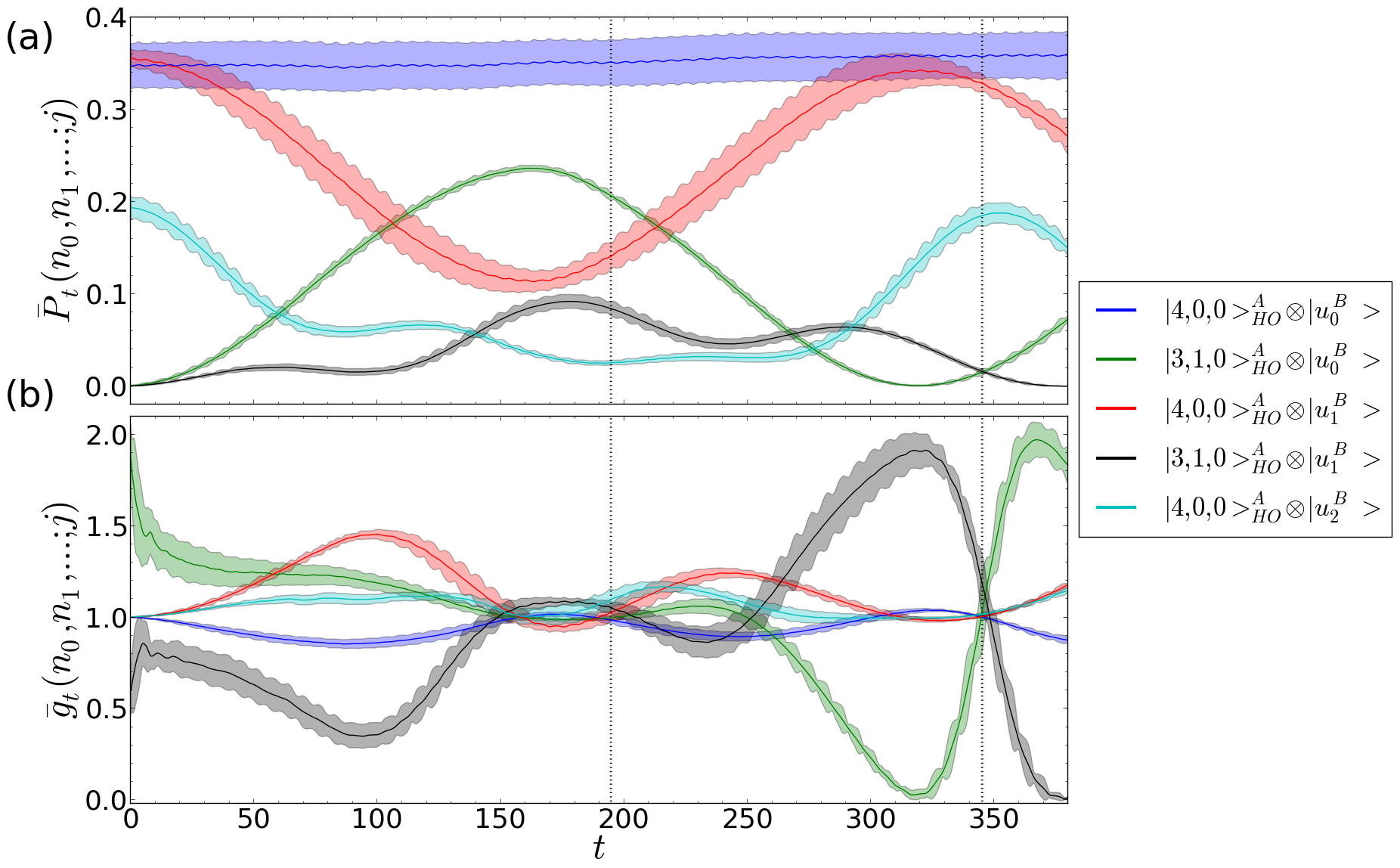}
\caption[Caption_1]{Joint probability $\bar P_t(n_0,n_1,...;j)$ (a) and
correlation
measure $\bar g_t(n_0,n_1,...;j)$ (b) for selected configurations of the
bipartite
system. The dashed vertical line at $t\approx195$ ($t\approx345$) indicates
the instant of the $\bar\Delta^B(t)$ minimum (maximum) in the considered 
time-interval.
The shaded areas indicate the standard deviation from the respective locally
time-averaged quantity.
Parameters: $N_A=4$, $\alpha=1$, $d=1.5$. All quantities shown in HO units of
$\hat H_B$.}
\label{fig:fock_correlations}
\end{figure}
In fig. \ref{fig:fock_correlations}, we present the evolution of both the
joint probability $\bar P_t(n_0,n_1,...;j)$ and the Fock space correlation
measure
$\bar g_t(n_0,n_1,...;j)$ for $N_A=4$ and selected configurations, which have a
significant contribution to the total wave function. We
emphasize that we do not intend to present a comprehensive overview over all
energy transfer channels but rather concentrate on some important ones.

As expected, the considered excitations turn out to
be uncorrelated whenever $\bar S_{\rm vN}(t)$ attains a deep minimum, which is
partly
related to extremal values of $\bar\Delta^B(t)$ as discussed in sect.
\ref{ssec:inter_spec_corr}.
In between those instants, the subsystem configurations
$|N_A,0,...\rangle^{A}_{\rm
HO}$, $|u^B_0\rangle$ are found slightly anti-bunched, while the occupation of
the
dynamically decoupled
one-dimensional $E_{\rm SP}(0)$ subspace stays rather constant.
In contrast to this, the
configurations of the $E_{\rm SP}(1)$ subspace feature more involved
correlations: Whereas the configurations $|N_A,0,...\rangle^{A}_{\rm
HO}$, $|u^B_1\rangle$ bunch, their counterparts $|N_A-1,1,0,...\rangle^{A}_{\rm
HO}$, $|u^B_0\rangle$ are measured with enhanced (reduced) probability when the
$B$ atom effectively donates (accepts) energy. The reverse statements hold
for the  configurations $|N_A-1,1,0,...\rangle^{A}_{\rm
HO}$, $|u^B_1\rangle$ belonging to the $E_{\rm SP}(2)$ subspace. We conclude
that, 
depending on the direction of the energy transfer,
inter-species correlations stochastically favour or disfavour
certain subsystem configurations and thereby energy transfer channels.

\section{Conclusions and outlook}\label{sec:concl}

In this work, we have analysed the correlated energy transfer between a single
atom and a finite, interacting bosonic environment mediated by a
spatio-temporally localized coupling, focusing on an initial excess energy of
the single
atom 
just above the environment excitation gap. By {\it ab-initio} 
simulations of the whole system, we have shown that the energy transfer
happens the faster the larger the number of environmental degrees of freedom
$N_A$ is, which is explained classically by an increased number of collision
partners and quantum-mechanically by a with $N_A$ increasing level
splitting of the
degenerate two first excited states of the uncoupled system. At the same time,
the energy transfer
between the subsystems becomes less complete when increasing $N_A$ so that the
excess energy distribution oscillates around quite a balanced one
for larger $N_A$, 
which might be regarded as a precursor of temporal
relaxation induced by
dephasing and has important 
consequences for inter-species correlations: While correlations
are analytically shown to be suppressed whenever the excess
energy distribution is maximally imbalanced, more balanced distributions allow
for stronger correlations, which are dynamically established indeed. By 
inspecting the energy distribution among the natural orbitals 
of the $B$ atom as well as the whole $A$ species,
incoherent transfer processes being intimately connected to inter-species
correlations are unravelled. While donating (accepting) energy, each subsystem
simultaneously shuffles a significant fraction of its energy into
the second dominant natural orbital and back. These incoherent processes
accelerate
the early energy donation of the $B$ atom and might be experimentally
detectable by correlating the outcomes of single shot energy
measurements upon both subsystems. Moreover, by analysing correlations
between 
subsystem excitations, we show 
that inter-species
correlation (dis-)favour certain energy transfer channels depending on the
instantaneous direction of transfer.

The correlated energy transfer dynamics manifests itself
in the subsystem dynamics as follows: The $B$ atoms experiences
a temporal loss of coherence in terms of short time periods during
which the Husimi function undergoes significant shape changes.
These deviations from a coherent state are more pronounced for larger $N_A$.
The bosonic environment features depletion oscillations 
with a maximal depletion decreasing faster with increasing $N_A$ than its upper,
energetically allowed bound $\propto 1/N_A$, which is derived for negligible
intra-species interaction. In the considered low-excess energy regime, the
excitations of the environment induced by the coupling to the $B$ atom mainly
consist of singlet and delayed doublet excitations, whose delay is explained 
by a temporally coarse-grained Schr\"odinger equation.

To round off this work, we have
made a proposal for an experimental
realization of this open quantum system problem based on
a combination of an optical dipole trap with a magnetic field gradient,
which allows for a species-selective trapping of selected ${}^{87}$Rb hyperfine
states.

As an outlook, our work on systems with spatio-temporally
localized couplings allows various interesting extensions: (i) Having understood
the low-excess energy regime, a natural step consists in a careful increase of
the excess energy in order to systematically open more and more energy transfer
channels. (ii) It is important to address the questions whether the excess
energy distribution among the subsystems temporally relaxes due to
dephasing and whether inter-species correlations are persistent when
significantly increasing $N_A$ to values $\mathcal{O}(100)$. However, 
such investigations
will require to either separate the traps further or switch off / rescale
the intra-species
interaction strength in order to keep the initial inter-species overlap
vanishing.
(iii) The demonstrated tunability of the inter-species coupling matrix elements
(\ref{eq:v_ijpq}) in terms of the trap separation $R$ should be exploited  to
control the correlated energy transfer.

\section*{Achnowledgements}
The authors would like to thank Oriol Vendrell, Hans-Dieter Meyer, Walter 
Strunz, Lushuai Cao and Antonio Negretti for inspiring discussions, Christof
Weitenberg for valuable support
concerning the experimental realizability of this system as well as Maxim
Efremov
for his expertise in phase-space analysis. S.K. acknowledges a scholarship of
the Studienstiftung des deutschen Volkes and P.S. acknowledges funding by the
Deutsche Forschungsgemeinschaft in the framework of the SFB 925 ``Light induced
dynamics and control of correlated quantum systems''.

\appendix
\section{ML-MCTDHB method}\label{app:method}
Rather than considering the reduced dynamics for the single atom in terms
of some effective master equation
\cite{Breuer_theory_open_quantum_systems}, we simulate here the correlated
quantum
dynamics of the total system by means of the 
recently developed {\it ab-initio} ML-MCTDHB method
\cite{kronke_non-equilibrium_2013,cao_multi-layer_2013}. Our method is based on
three key ideas: (i) The total many-body wave function is expanded with respect
to a time-dependent, optimally with the system co-moving many-body basis, which
allows to 
obtain converged results with a significantly reduced number of basis states
compared to expansions relying on a time-independent basis. (ii) The
symmetry of the bosonic species is explicitly employed. (iii) The multi-layer
ansatz for the total wave
function is based on a coarse-graining cascade, which enables to adapt the
ansatz to intra- and inter-species correlations and makes
ML-MCTDHB to be a highly flexible, versatile tool for simulating the full
dynamics of bipartite systems. Thereby, ML-MCTDHB incorporates both
the coarse-graining concept of the Multi-Layer Multi-Configuration
Time-Dependent Hartree Method 
\cite{Multilayer_formulation_Wang_Thoss_2003,Multilayer_multiconfig_Manthe_2008,
ML_implementation_Vendrell_Meyer_JChemPhys_2011}
for distinguishable degrees of freedom and
the explicit consideration of the particle exchange symmetry in the 
Multi-Configuration Time-Dependent Hartree Method for Bosons
\cite{role_of_excited_states_in_splitting_of_BEC_by_time_dep_barrier_Steltsov_PRL2007,
MCTDHB_many_body_dynacmics_of_bosonic_systems_phys_rev_a_Alon_Streltsov_Cederbaum}
(cf. also \cite{MLMCTDH_in_2nd_quantization_representation_Wang_Thoss_Chem_Phys_2009}
in this context).
Compared to the direct extension of the latter method to mixtures
\cite{mctdh_for_mixtures_of_2types_of_identical_particles_PRA2007}, ML-MCTDHB 
can approach larger system sizes because of its adaptability to system-specific 
inter-species
correlations.

Here, we summarize the ML-MCTDHB ansatz concretised for a binary bosonic
mixture with particle numbers $N_A$ and $N_B$. Firstly, we assign
$m_\sigma$ time-dependent single-particle basis functions
$|\phi^\sigma_i(t)\rangle$,
 $i=1,...,m_\sigma$ to the species $\sigma=A,B$. These single-particle basis
functions
are then raised to a time-dependent $N_\sigma$-body basis by constructing
bosonic number states $|n_1,...n_{m_\sigma}\rangle_t^\sigma$ describing 
a fully symmetrized Hartree product with $n_i$ bosons of the $\sigma$ species
occupying
the state $|\phi^\sigma_i(t)\rangle$. Instead of considering all 
$(N_\sigma+m_\sigma-1)!/[N_\sigma!(m_\sigma-1)!]$ number state configurations
on an equal footing, we consider only a reduced number of $M$ time-dependent
species basis functions:
\begin{equation}\label{eq:species_basis}
 |\psi^\sigma_j(t)\rangle = \sum_{n_1,...,n_{m_\sigma}|N_\sigma}
 C^\sigma_{j;n_1,...,n_{m_\sigma}}(t)
 |n_1,...n_{m_\sigma}\rangle_t^\sigma, \quad
\end{equation}
where $j=1,...,M$ and ``$...|N_\sigma$'' indicates that all
occupation numbers must add up to $N_\sigma$. These ``coarse-grained'' 
species basis functions are finally used in order to expand the total wave
function of the binary mixture:
\begin{equation}\label{eq:toplayer_exp}
 |\Psi_t\rangle=\sum_{i_A,i_B=1}^{M}A_{i_A i_B}(t)
 |\psi^A_{i_A}(t)\rangle\otimes|\psi^B_{i_B}(t)\rangle.
\end{equation}
Having specified the ML-MCTDHB wave function ansatz, one employs a variational
principle such as the Dirac-Frenkel \cite{dirac_variational,frenkel_variational}
or equivalently the McLachlan variational principle
\cite{mc_lachlan_1963} for deriving 
the equations of motion both for the expansion coefficients
$A_{i_A i_B}$ and the basis functions on the species layer, i.e.
$C^\sigma_{j;n_1,...,n_{m_\sigma}}$, as well as on the particle layer, i.e.
$|\phi^\sigma_i\rangle$. Relying on a variational principle, these equations of
motion ensure that the basis states move
in a variationally optimally way so that at each instant in time one obtains
an optimal representation of the total wave function for given numbers of 
species basis functions $M$ and single-particle basis function $m_\sigma$.
The resulting set of coupled ordinary differential equations and non-linear
partial differential equations, however, is too cumbersome to be
discussed here and we refer to
\cite{kronke_non-equilibrium_2013,cao_multi-layer_2013} for the details. We 
note that the ML-MCTDHB method also allows for calculating low lying excited
states by means of the improved relaxation algorithm
\cite{cao_multi-layer_2013,
quantum_molecular_dynamics_using_mctdh_Meyer_Worth_2002}.

One key feature of the ML-MCTDHB ansatz lies in its flexibility (iii): It covers
both the mean-field limit $M=m_\sigma=1$ and the numerical exact limit. The
latter is achieved by considering as many single-particle particle basis
functions as one employs time-independent basis states, e.g. grid points, in
order to represent them and choosing
$M={(N_\sigma+m_\sigma-1)!/[N_\sigma!(m_\sigma-1)!]}$. Moreover, in the presence
of quite some intra-species but weak inter-species correlations, as it is often
the case for system-environment problems, one might get converged results
with $M$ being much smaller than
${(N_\sigma+m_\sigma-1)!/[N_\sigma!(m_\sigma-1)!]}$. 
In our particular situation with $N_B=1$, one has to choose $m_B=M$. So
we have to ensure convergence with respect to numbers of grid points $n$
and time-dependent basis functions, i.e. $m_A$ and $M$.
By setting the number of time-dependent
optimized species layer basis functions $M=1$ while still
bringing the simulation to convergence with respect to the number $m_A$ of
time-dependent optimized $A$ boson single-particle basis functions, 
the ML-MCTDHB equations of motion give a variationally
optimized solution for the many-body problem under the constraint of vanishing
inter-species correlations, which can illuminate the role of inter-species
correlations as discussed in sect. \ref{ssec:E_NORBs}.

  \section{Initial state preparation and convergence}\label{app:init_converg}
In order to prepare our initial state, we first shift the harmonic trap
for the single atom to the right by replacing 
$(\hat x^B)^2/2$ by $(\hat x^B-d)^2/2$ in $\hat H_B$ and propagate an initial guess for
(\ref{eq:toplayer_exp}) with the ML-MCTDHB equations of motion in imaginary
time. Thereby,
we relax the binary mixture to its many-body ground state. Due to the almost
vanishingly small spatial overlap of the two species, the result of this 
procedure is a many-body wave function factorizing to good approximation into
the ground state of $N_A$ interacting bosons in their harmonic trap and 
the single atom being in the coherent state $|z=d/\sqrt{2}\rangle$. Afterwards,
we propagate this state with the unshifted Hamiltonian (\ref{eq:hamilt_sep}).

For representing the time-dependent single-particle basis
functions, we employ a harmonic
oscillator discrete variable representation
\cite{dvr_and_their_utilization_Light_Carrington_2000,MCTDH_BJMW2000} with
$n=401$ grid points. 
Concerning the size of the time-dependent basis, we ensured convergence 
by checking that incrementing $m_A$
or $M$ does at most slightly quantitatively
affect physical observables such as energy and the natural populations on the
species and particle layers.
For our purposes, $m_A = 3$ and $M = 4$ turned out to lead to
convergence for $N_A < 10$ and $d=1.5$. For $N_A=10$, a $(m,M)=(3,3)$
simulation agree well with a  $(3,4)$ run up to $t\sim 200$. Afterwards,
the natural populations of the $(3,4)$ simulation still indicate
convergence, yet an explicit convergence check for such long-time propagations
by further increasing $m$ and $M$ 
is computationally prohibitive. Due to an decrease of the system
time-scale $T_{\rm cycle}$ with $N_A$, it is very natural that high precision
results until $t=700$ are much harder to obtain for $N_A=10$ compared to smaller
boson numbers and so we have to work with variationally optimal results of somewhat
lower accuracy. Similarly, simulations with larger excess energies are more
demanding: For $N_A = 2$ and $d=2.5$, we have checked the plotted $(3,4)$
results against $(5,5)$ simulations, indicating that the accuracy is somewhat
lower compared to the case $d=1.5$ but sufficiently high for the considered
observables.

\section{Bound on the $A$ species depletion}\label{app:proof_condensation}
Neglecting the intra-species interaction, we show here that the depletion of
the $A$ species vanishes as $1/N_A$ in the large $N_A$ limit as a consequence
of the fixed excess energy and $\hat H_A$ 
featuring a gapped excitation spectrum.
Setting $g_A=0$ and assuming the initial inter-species interaction
energy to be negligible $\langle \hat H_{AB}\rangle_0\ll\langle
\hat H\rangle_0$ (which is the case to a very good approximation in fact), 
energy conservation implies:
\begin{equation}\nonumber
 N_A E^A_0 + E^B_0+\varepsilon \approx \langle \hat H_{A}\rangle_t+
  \langle \hat H_{B}\rangle_t +\langle \hat H_{AB}\rangle_t
  \geq \langle \hat H_{A}\rangle_t + E^B_0,
\end{equation}
given the excess energy $\varepsilon=d^2/2$. Since $\hat H_A$ contains only the
single-particle Hamiltonians $\hat h^A_i={[({\hat p^A_i})^2 +
(\hat x^A_i + R)^2]/2}$ by assumption, we may express $\langle \hat H_{A}\rangle(t)$
as a function(al) of the $\hat\rho^A_t$  NPs / NOs sorted in
increasing sequence with respect to their instantaneous excess energy
$\tilde\varepsilon^A_i(t)=
\langle\tilde\varphi^A_i(t)|\hat h^A_1|\tilde\varphi^A_i(t)\rangle - E^A_0$ being
indicated by a tilde and obtain:
\begin{equation}\label{eq:Anorb_excess_inequal}
\tilde\varepsilon^A_1(t)+\sum_{i=2}^{m_A}
\tilde\lambda_i^A(t)[\tilde\varepsilon^A_i(t)-\tilde\varepsilon^A_1(t)]\lesssim
\frac{ \varepsilon}{N_A }.
\end{equation}
From now on, we omit the tilde and the time-dependencies in the notation.
The excess energy  $\varepsilon^A_1$ of the energetically lowest lying NO is
thus bounded from above by 
$\varepsilon/N_A$ as a consequence of $\langle \hat H_{A}\rangle$ being
extensive. Assuming now $N_A\gg \varepsilon/(E^A_1-E^A_0)$, similar arguments
concerning the orthonormality of the NOs as explicated in appendix
\ref{app:proof_disentanglement} apply, showing $\varepsilon^A_i\gtrsim
E^A_1-E^A_0 = 1 \gg \varepsilon/N_A$ for $i>1$. Finally, we may employ
(\ref{eq:Anorb_excess_inequal}) in order to prove:
\begin{equation}
\lambda_i^A\lesssim\frac{\varepsilon/N_A-\varepsilon^A_1}{E^A_1-E^A_0}
\lesssim\frac{d^2}{2N_A},\quad i>1.
\end{equation}
Thus, the $A$ species becomes condensed as $N_A\rightarrow \infty$.

\section{Absence of correlations for strongly
imbalanced excess energy distribution between
subsystems}\label{app:proof_disentanglement}
In this appendix, we sketch the proof for the fact that the excess
energy $\varepsilon^B_t$ of the $B$ atom
being much smaller than its excitation gap $(E^B_1-E^B_0)$ implies the
absence
of significant inter-species correlations: Reordering the NOs in an increasing
sequence with respect to their energy content $\Sigma^B_i(t)$, one obtains 
the representation:
\begin{equation}
 \hat\rho^B_t = \Big(1-\sum_{j=2}^{m_B} \tilde \lambda_j^B(t) \Big)\,
|\tilde\varphi^\sigma_1(t)\rangle\!\langle\tilde\varphi^\sigma_1(t)|
+\sum_{i=2}^{m_B}\tilde
\lambda_i^B(t)|\tilde\varphi^\sigma_i(t)\rangle\!\langle\tilde\varphi^\sigma_i(
t )|.
\end{equation}
From now on, we omit the tilde denoting this particular ordering as well as
all time-dependencies in the notation. With this
expression, the excess
energy $\varepsilon^B$ of $B$ can be decomposed into 
the excess energies
of the NOs obeying the ordering $\varepsilon^B_{i+1}\geq \varepsilon^B_i$:
\begin{equation}
 \varepsilon^B = \varepsilon_1^B+\sum_{i=2}^{m_B}
\lambda_i^B(\varepsilon_i^B-\varepsilon_1^B),
\end{equation}
which leads to $\varepsilon_1^B\leq \varepsilon^B$. Furthermore, we will show
$\varepsilon_i^B > \varepsilon_1^B$ for all $i\geq 2$ below implying an upper
bound for the $i$-th NP:
\begin{equation}\label{eq:npop_upper_bound}
\lambda_i^B \leq
\frac{\varepsilon^B-\varepsilon_1^B}{\varepsilon_i^B-\varepsilon_1^B},\quad
i\geq
2.
\end{equation}
Expanding $|\varphi^B_i\rangle$
in terms of the HO eigenstates $|u^B_0\rangle$, $|u^B_1\rangle$ and a normalized
vector $|v^i_\bot\rangle$ orthogonal to the former ones,
\begin{equation}
 |\varphi^B_i\rangle = c^i_0|u^B_0\rangle+c^i_1|u^B_1\rangle
		      + c^i_\bot|v^i_\bot\rangle,
\end{equation}
we find $\varepsilon_i^B=|c^i_1|^2(E^B_1-E^B_0)+|c^i_\bot|^2(E^i_\bot-E^B_0)$
where $E^i_\bot=\langle v^i_\bot|\hat H_B|v^i_\bot\rangle\geq E^B_2$ 
according to the
Ritz variational principle. This identity can be employed to see that 
$\varepsilon_1^B\leq \varepsilon^B \ll E^B_1-E^B_0$ implies $|c^1_1|^2,
|c^1_\bot|^2
\ll 1$ and thus $|c^1_0|^2=\mathcal{O}(1)$. The orthonormality of the NOs
prevents any other NO to have $|c^i_0|^2=\mathcal{O}(1)$ such that
$\varepsilon_i^B\geq \mathcal{O}(E^B_1-E^B_0)$ for $i\geq 2$ proving the
anticipated inequality 
$\varepsilon_i^B > \varepsilon_1^B$ for $i\geq 2$. The resulting inequality
$\varepsilon_1^B\leq \varepsilon^B \ll \varepsilon_i^B$ for $i\geq 2$ and the
bound 
(\ref{eq:npop_upper_bound}) finally imply $\lambda_i^B\ll 1$ for $i\geq 2$.
The existence of only a single NO with excess energy of
$\mathcal{O}(\varepsilon^B)\ll(E^B_1-E^B_0)$ thus results in the absence of
inter-species correlations.

\section{Persistence of inter-species correlations under $N_A$ {\bf increase}}
\label{app:example_persist_corr}
Neglecting complications due to the intra-species interaction,
we have shown in \ref{app:proof_condensation} that 
the $A$ species becomes condensed as $N_A\rightarrow \infty$. This might
seem to contradict the apparently persistent inter-species correlations
discussed in sect. \ref{ssec:inter_spec_corr} since $\lambda_1^A=1$ 
would necessarily
imply $\lambda_1^B=1$ given that the total system is in a pure state. In order
to disprove this seeming contradiction, we provide a minimal example
illustrating that inter-species correlations can survive the $N_A\rightarrow
\infty$ limit:
Let us assume that the state of the $A$ species is given by:
\begin{equation}\label{eq:min_example_species}
\hat \eta^A\simeq \frac{1}{2}\left( 
\begin{array}{cc}
1+a & b \\
b^* & 1- a
\end{array}\right),
\end{equation}
where the corresponding basis vectors are given by
$(1,0)^T\simeq|N_A,0\rangle^{A}_{\rm HO}$ and 
$(0,1)^T\simeq|N_A-1,1\rangle^{A}_{\rm HO}$.
Equation
(\ref{eq:min_example_species}) defines a density
operator for all $a\in\mathbb{R}$, $b\in\mathbb{C}$ such that the Bloch vector
norm obeys $|{\bf n}|^2\equiv a^2+|b|^2\leq 1$ and is consistent with the
typical
energies of the $A$ species considered in this work. The corresponding NPs
are given by $\lambda^B_{1/2}=(1\pm|{\bf n}|)/2$ and, thus, any inter-species
entanglement entropy $S_{\rm vN}\in[0,\ln 2]$ can be realized by appropriately
choosing $|{\bf n}|$. For the above state $\hat \eta^A$, the
depletion of the $A$ species,
\begin{equation}
 1-\lambda^A_1=\frac{2(1-a)-|b|^2}{4N_A}+\mathcal{O}(N_A^{-2}),
\end{equation}
vanishes in the large $N_A$ limit independently of the inter-species
correlations.

\section*{References}
\bibliography{references_without_title}

\begin{thebibliography}{10}

\bibitem{quantum_diss_systems_Weiss}
Weiss U.
\newblock {\em {Q}uantum {D}issipative {S}ystems}, volume~13 of {\em Series in
  Modern Condensed Matter Physics}.
\newblock World Scientific, 4th edition, 2012.

\bibitem{open_sys_approach_qo}
Carmichael H.
\newblock {\em {A}n {O}pen {S}ystems {A}pproach to {Q}uantum {O}ptics},
  volume~18 of {\em Lecture Notes in Physics Monographs}.
\newblock Springer Berlin Heidelberg, 1993.

\bibitem{vendrell_proton_2008}
Vendrell O and Meyer H-D.
\newblock {\em Phys. Chem. Chem. Phys.}, 10:4692, 2008.

\bibitem{ritschel_efficient_2011}
Ritschel G, Roden J, Strunz~W T, and Eisfeld A.
\newblock {\em New J. Phys.}, 13:113034, 2011.

\bibitem{roden_probability_2015}
Roden J~J J and Whaley~K B.
\newblock {\em {arXiv}:1501.06090 [quant-ph]}, 2015.

\bibitem{Breuer_theory_open_quantum_systems}
Breuer H-P and Petruccione F.
\newblock {\em {T}he {T}heory of {O}pen {Q}uantum {S}ystems}.
\newblock Oxford University Press, 2002.

\bibitem{many_body_physics_ultrac_atoms_zwerger_dalibard_bloch}
Bloch I, Dalibard J, and Zwerger W.
\newblock {\em Rev. Mod. Phys.}, 80:885, 2008.

\bibitem{muller_engineered_2012}
M\"uller M, Diehl S, Pupillo G, and Zoller P.
\newblock {E}ngineered {O}pen {S}ystems and {Q}uantum {S}imulations with
  {A}toms and {I}ons.
\newblock In P~Berman, E~Arimondo, and C~Lin, editors, {\em Advances In Atomic,
  Molecular, and Optical Physics}, volume~61, pages 1--80. Academic Press,
  2012.

\bibitem{barreiro_open-system_2011}
Barreiro~J T, M\"uller M, Schindler P, Nigg D, Monz T, Chwalla M, Hennrich M,
  Roos~C F, Zoller P, and Blatt R.
\newblock {\em Nature}, 470:486, 2011.

\bibitem{schindler_quantum_2013}
Schindler P, M\"uller M, Nigg D, Barreiro~J T, Martinez~E A, Hennrich M, Monz
  T, Diehl S, Zoller P, and Blatt R.
\newblock {\em Nat. Phys.}, 9:361, 2013.

\bibitem{scelle_motional_2013}
Scelle R, Rentrop T, Trautmann A, Schuster T, and Oberthaler~M K.
\newblock {\em Phys. Rev. Lett.}, 111:070401, 2013.

\bibitem{bakr_quantum_2009}
Bakr~W S, Gillen~J I, Peng A, F\"olling S, and Greiner M.
\newblock {\em Nature}, 462:74, 2009.

\bibitem{sherson_single-atom-resolved_2010}
Sherson~J F, Weitenberg C, Endres M, Cheneau M, Bloch I, and Kuhr S.
\newblock {\em Nature}, 467:68, 2010.

\bibitem{weitenberg_single-spin_2011}
Weitenberg C, Endres M, Sherson~J F, Cheneau M, Schau{\ss} P, Fukuhara T, Bloch
  I, and S~Kuhr.
\newblock {\em Nature}, 471:319, 2011.

\bibitem{deterministic_preparation_of_tunable_few-fermion_system_selim_Science2011}
Serwane F, Z\"urn G, Lompe T, Ottenstein~T B, Wenz~A N, and Jochim S.
\newblock {\em Science}, 332:336, 2011.

\bibitem{spethmann_dynamics_2012}
Spethmann N, Kindermann F, John S, Weber C, Meschede D, and Widera A.
\newblock {\em Phys. Rev. Lett.}, 109:235301, 2012.

\bibitem{palzer_quantum_2009}
Palzer S, Zipkes C, Sias C, and K\"ohl M.
\newblock {\em Phys. Rev. Lett.}, 103:150601, 2009.

\bibitem{johnson_impurity_2011}
Johnson~T H, Clark~S R, Bruderer M, and Jaksch D.
\newblock {\em Phys. Rev. A}, 84:023617, 2011.

\bibitem{fukuhara_quantum_2013}
Fukuhara T, Kantian A, Endres M, Cheneau M, Schau{\ss} P, Hild S, Bellem D,
  Schollw\"ock U, Giamarchi T, Gross C, Bloch I, and Kuhr S.
\newblock {\em Nat. Phys.}, 9:235, 2013.

\bibitem{catani_quantum_2012}
Catani J, Lamporesi G, Naik D, Gring M, Inguscio M, Minardi F, Kantian A, and
  Giamarchi T.
\newblock {\em Phys. Rev. A}, 85:023623, 2012.

\bibitem{peotta_quantum_2013-1}
Peotta S, Rossini D, Polini M, Minardi F, and Fazio R.
\newblock {\em Phys. Rev. Lett.}, 110:015302, 2013.

\bibitem{barontini_controlling_2013}
Barontini G, Labouvie R, Stubenrauch F, Vogler A, Guarrera V, and Ott H.
\newblock {\em Phys. Rev. Lett.}, 110:035302, 2013.

\bibitem{diehl_quantum_2008}
Diehl S, Micheli A, Kantian A, Kraus B, B\"uchler~H P, and Zoller P.
\newblock {\em Nat. Phys.}, 4:878, 2008.

\bibitem{verstraete_quantum_2009}
Verstraete F, Wolf~M M, and Cirac~J I.
\newblock {\em Nat. Phys.}, 5:633, 2009.

\bibitem{pastawski_quantum_2011}
Pastawski F, Clemente L, and Cirac~J I.
\newblock {\em Phys. Rev. A}, 83:012304, 2011.

\bibitem{harter_single_2012}
H\"arter A, Kr\"ukow A, Brunner A, Schnitzler W, Schmid S, and Denschlag~J H.
\newblock {\em Phys. Rev. Lett.}, 109:123201, 2012.

\bibitem{coll_gate_Zoller}
Jaksch D, Briegel H-J, Cirac~J I, Gardiner~C W, and Zoller P.
\newblock {\em Phys. Rev. Lett.}, 82:1975, 1999.

\bibitem{mack_dynamics_2002}
Mack H and Freyberger M.
\newblock {\em Phys. Rev. A}, 66:042113, 2002.

\bibitem{hutton_entangle_by_scattering}
Harshman~N L and Hutton G.
\newblock {\em Phys. Rev. A}, 77:042310, 2008.

\bibitem{benedict_time_2012}
Benedict~M G, Kov\'{a}ics J, and Czirj\'{a}ik A.
\newblock {\em J. Phys. A: Math. Theor.}, 45:085304, 2012.

\bibitem{law_entanglement_2004}
Law~C K.
\newblock {\em Phys. Rev. A}, 70:062311, 2004.

\bibitem{sowinski_dynamics_2010}
Sowi\'{n}ski T, Brewczyk M, Gajda M, and Rz\k{a}\.{z}ewski K.
\newblock {\em Phys. Rev. A}, 82:053631, 2010.

\bibitem{Gardiner_PRA2013}
Holdaway D~I H, Weiss C, and Gardiner~S A.
\newblock {\em Phys. Rev. A}, 87:043632, 2013.

\bibitem{kinoshita_quantum_2006}
Kinoshita T, Wenger T, and Weiss~D S.
\newblock {\em Nature}, 440:900, 2006.

\bibitem{ganahl_quantum_2013}
Ganahl M, Haque M, and Evertz~H G.
\newblock {\em {arXiv}:1302.2667}, 2013.

\bibitem{franzosi_newtons_2014}
Franzosi R and Vaia R.
\newblock {\em J. Phys. B: At. Mol. Opt. Phys.}, 47:095303, 2014.

\bibitem{kronke_non-equilibrium_2013}
Kr\"onke S, Cao L, Vendrell O, and Schmelcher P.
\newblock {\em New J. Phys.}, 15:063018, 2013.

\bibitem{cao_multi-layer_2013}
Cao L, Kr\"onke S, Vendrell O, and Schmelcher P.
\newblock {\em J. Chem. Phys.}, 139:134103, 2013.

\bibitem{sys_env_corr_non_mark_ansgar}
Pernice A, Helm J, and Strunz~W T.
\newblock {\em J. Phys. B: At. Mol. Opt. Phys.}, 45:154005, 2012.

\bibitem{env_collec_reaction_coord}
Iles-Smith J, Lambert N, and Nazir A.
\newblock {\em Phys. Rev. A}, 90:032114, 2014.

\bibitem{Pethick2002}
Pethick~C J and Smith H.
\newblock {\em {B}ose-{E}instein {C}ondensates in {D}ilute {G}ases}.
\newblock Cambridge University Press, 2nd edition, 2008.

\bibitem{Olshanii_PRL98_quasi_1d_scattering}
M.~Olshanii.
\newblock {\em Phys. Rev. Lett.}, 81:938, 1998.

\bibitem{Pitaevskii_Stringari_Bose-Einstein_Condensation2003}
Stringari S and Pitaevskii~L P.
\newblock {\em {B}ose-{E}instein {C}ondensation}.
\newblock Oxford University Press, 2003.

\bibitem{krych_displaced_traps}
Krych M and Idziaszek Z.
\newblock {\em Phys. Rev. A}, 80:022710, 2009.

\bibitem{busch_nonlocality}
Fogarty T, Busch T, Goold J, and Paternostro M.
\newblock {\em New J. Phys.}, 13:023016, 2011.

\bibitem{fermionizations_of_2_distinguishable_fermions_ZuernSelim_PRL2012}
Z\"urn G, Serwane F, Lompe T, Wenz~A N, Ries~M G, Bohn~J E, and Jochim S.
\newblock {\em Phys. Rev. Lett.}, 108:075303, 2012.

\bibitem{BergFriedlander:2008}
Berg~E van den and Friedlander~M P.
\newblock {\em SIAM J. Sci. Comput.}, 31:890, 2008.

\bibitem{spgl1:2007}
Berg~E van den and Friedlander~M P.
\newblock {SPGL1}: {A} solver for large-scale sparse reconstruction, June 2007.
\newblock http://www.cs.ubc.ca/labs/scl/spgl1.

\bibitem{loewdin_norb55}
L\"owdin P-O.
\newblock {\em Phys. Rev.}, 97:1474, 1955.

\bibitem{Onsager_Penrose_BEC_liquid_He_PR_1956}
Penrose O and Onsager L.
\newblock {\em Phys. Rev.}, 104:576, 1956.

\bibitem{glauber_quantum_1963}
Glauber~R J.
\newblock {\em Phys. Rev.}, 130:2529, 1963.

\bibitem{Multilayer_formulation_Wang_Thoss_2003}
Wang H and Thoss M.
\newblock {\em J. Chem. Phys.}, 119:1289, 2003.

\bibitem{Multilayer_multiconfig_Manthe_2008}
Manthe U.
\newblock {\em J. Chem. Phys.}, 128:164116, 2008.

\bibitem{ML_implementation_Vendrell_Meyer_JChemPhys_2011}
Vendrell O and Meyer H-D.
\newblock {\em J. Chem. Phys.}, 134:044135, 2011.

\bibitem{role_of_excited_states_in_splitting_of_BEC_by_time_dep_barrier_Steltsov_PRL2007}
Streltsov~A I, Alon~O E, and Cederbaum~L S.
\newblock {\em Phys. Rev. Lett.}, 99:030402, 2007.

\bibitem{MCTDHB_many_body_dynacmics_of_bosonic_systems_phys_rev_a_Alon_Streltsov_Cederbaum}
Alon~O E, Streltsov~A I, and Cederbaum~L S.
\newblock {\em Phys. Rev. A}, 77:033613, 2008.

\bibitem{MLMCTDH_in_2nd_quantization_representation_Wang_Thoss_Chem_Phys_2009}
Wang H and Thoss M.
\newblock {\em J. Chem. Phys.}, 131:024114, 2009.

\bibitem{mctdh_for_mixtures_of_2types_of_identical_particles_PRA2007}
Alon~O E, Streltsov~A I, and Cederbaum~L S.
\newblock {\em Phys. Rev. A}, 76:062501, 2007.

\bibitem{dirac_variational}
Dirac P~A M.
\newblock {\em Proc. Cambridge Philos. Soc.}, 26:376, 1930.

\bibitem{frenkel_variational}
Frenkel J.
\newblock {\em {W}ave {M}echanics}.
\newblock Clarendon Press, Oxford, 1934.

\bibitem{mc_lachlan_1963}
McLachlan~A D.
\newblock {\em Mol. Phys.}, 8:39, 1963.

\bibitem{quantum_molecular_dynamics_using_mctdh_Meyer_Worth_2002}
Meyer H-D and Worth~G A.
\newblock {\em Theor. Chem. Acc.}, 109:251, 2003.

\bibitem{dvr_and_their_utilization_Light_Carrington_2000}
Light~J C and Carrington T.
\newblock {\em Adv. Chem. Phys.}, 114:263, 2000.

\bibitem{MCTDH_BJMW2000}
Beck~M H, J\"ackle A, Worth~G A, and Meyer H-D.
\newblock {\em Phys. Rep.}, 324:1, 2000.

\end{thebibliography}
\bibliographystyle{unsrt}

\end{document}